\documentclass[11pt,preprintnumbers,pra,aps]{revtex4}
\usepackage{amssymb,amsmath,amsbsy,amsfonts,amsthm,mathtools,bm}
\usepackage[breaklinks,colorlinks,citecolor=blue,urlcolor=violet]{hyperref}
\usepackage[mediumqspace,mediumspace,amssymb,Gray]{SIunits}
\usepackage{dcolumn,longtable,rotating,tabularx,array,lscape,braket,url}

\usepackage{xcolor,soul,cancel,enumitem,ulem,verbatim,graphicx,epsfig}
\usepackage[T1]{fontenc}
\usepackage{natbib}
\usepackage[most]{tcolorbox}
\usepackage{multirow}

\usepackage[caption=false,justification=justified]{subfig}
\begin{document}
\title{Proposal for Composite Quantum Electromagnetically Induced Transparency Heat Engine Coupled by a Nanomechanical Mirror}
\author{Rejjak Laskar\footnote{Email: laskarrejjak786@gmail.com}}
\affiliation
{\it
Department of Physics, Aliah University, IIA/27, Newtown, Kolkata 700160, India
}

\date{\today}
\begin{abstract}
This paper introduces a quantum heat engine model that utilizes an ultracold atomic gas coupled with a nanomechanical mirror. The mirror's vibration induces an opto-mechanical sideband in the control field, affecting the behavior of the cold gas and subsequently influencing the output radiation of the engine. The model incorporates mirror vibration while omitting cavity confinement, establishing a bridge between a multi-level atom-laser interacting system that plays with coherences and the mechanical vibration of the nanomechanical mirror, which jointly function as heat engines. Three distinct heat engine configurations are proposed: the first involves a vibration-free three-level $\Lambda$-type system, the second introduces nanomechanical vibration to the three-level $\Lambda$-type system, and the third constitutes a composite engine that combines the previous setups along with nanomechanical vibration. The spectral brightness of a three-level heat engine is diminished with mirror vibration, whereas for a composite heat engine, there is a slight enhancement in the brightness peak. However, the maximum brightness is attained when there is no vibration. Comparisons between the proposed model and an ideal system are made regarding entropy balance, adhering to the constraints of the second law of thermodynamics. The model observed that when subjected to mirror vibration, the proposed heat engines diverged from the characteristics expected in an ideal heat engine.
\end{abstract}
\maketitle
 \section{Introduction}
 
 A heat engine is a device that converts heat into work by absorbing heat from a hot bath and ejecting some of it into a cold bath with the efficiency bound by the second law of thermodynamics \cite{zemansky_heat_1981}. The primary examples of classical and quantum heat engines are a gas or steam turbine and a three-level laser, respectively. The present model specifically emphasizes the quantum engine instead of the classical one. The three-level laser is a perfect example of a continuous heat engine that combines quantum mechanics and thermodynamics on common ground, starting from the first principles \cite{doi:10.1146/annurev-physchem-040513-103724}. A three-level laser heat engine aims to convert population inversion, a quantum phenomenon, into light as its output. The study of quantum heat engines started several decades ago with Scovil \textit{et} \textit{al.} \cite{PhysRevLett.2.262} demonstrating that a three-level maser can function as a heat engine, limited by Carnot efficiency \cite{PhysRev.156.343}. Over time, the three-level system has been widely adopted as a model for quantum heat engines and refrigerators \cite{doi:10.1063/1.1776991, PhysRevE.49.3903, doi:10.1063/1.471453, doi:10.1080/09500340110090477, PhysRevLett.87.220601, HUMPHREY2005390, PhysRevA.74.063823, PhysRevLett.98.240601, scully2011quantum, Harbola_2012, PhysRevA.88.013842, PhysRevX.5.031044, PhysRevA.94.053859, PhysRevA.96.063806, PhysRevLett.119.050602, PhysRevA.103.062205, PhysRevResearch.2.043187}.

In particular, I mention the work of Haris \cite{PhysRevA.94.053859}, where a three-level $\Lambda$-type system is used as a heat engine under typical conditions of electromagnetically induced transparency (EIT) \cite{PhysRevLett.64.1107, PhysRevLett.66.2593, PhysRevA.52.2302}. In atomic systems, EIT is achieved through pump-probe spectroscopy, where the presence of a strong pump field makes the atomic system transparent to the probe field. Haris has used this advantage of EIT to develop a nontraditional quantum heat engine where two black bodies are used as photon reservoirs. Under the typical EIT condition, a strong pumping field is used with these two photon reservoirs to form a $\Lambda$-type system that emits radiation 64.9 times brighter than that predicted by Kirchhoff’s law. Based on Haris's theoretical work, Zou \textit{et} \textit{al.} \cite{PhysRevLett.119.050602} performed an experiment by using the cold $^{85}\textrm{Rb}$ atom in D$_{1}$-line. They have found that the brightness is about 9 times greater than that of the ambient pumping reservoir. In a Doppler-broadened medium, Zhang \textit{et} \textit{al.} \cite{PhysRevA.103.062205} extended Haris's work and showed that as the angular deviation of the emitted probe field increases, there is a decrease in brightness. The EIT-based quantum heat engines proposed by Harris and Zhang \textit{et al.} are compared with an ideal heat engine by checking the entropy balance.

In recent years, optomechanics \cite{aspelmeyer2012quantum} has emerged as a prominent field, bridging the interaction between light and mechanics to explore quantum physics' limits \cite{RevModPhys.86.1391, bawaj2015probing} and advancements in metrology \cite{4bc42494787a4610b87a9274d5ece0e5}. The quantum heat engine extracts work from the quantum system by exchanging heat with thermal reservoirs. In this regard, cavity optomechanics provides two advantages: radiation pressure helps to exchange energy between the cavity photons and mechanical phonons, and cavity damping couples the system with both reservoirs \cite{PhysRevLett.112.150602}. Polariton modes have been employed as the working substance for optomechanical heat engines, leading to various implementations \cite{PhysRevA.90.023819, PhysRevA.92.033854, PhysRevLett.115.223602, PhysRevLett.114.183602, Mari_2015}. As a result, there are two distinct approaches to handling engines at a small scale: optomechanical heat engines and three-level light-atom interacting systems. Combining these two approaches, particularly by employing a multi-level light-atom interacting system within the EIT framework as a heat engine associated with nanomechanical mirror vibration, introduces intriguing possibilities for exploration, such as

\begin{itemize}
      \item The effect of nanomechanical vibration on the emitted output light from the heat engine is intriguing, especially under typical electromagnetically induced transparency (EIT) conditions, where the emitted field is susceptible to quantum interference phenomena.

    \item  The vibration of the mirror can enhance the brightness of emitted photons under EIT conditions, making the system more attractive for observation. However, it is also possible that the vibration acts as a mere perturbation without any significant impact on the emitted photons.
    \item The most interesting aspect is studying how mirror vibration affects the entropy of the emitted radiation flux while adhering to the boundaries of the second law of thermodynamics.
\end{itemize}

In this paper, a novel scheme of a quantum heat engine is proposed, which incorporates nanomechanical mirror vibrations without cavity confinement. Unlike most optomechanical systems, the proposed model does not rely on cavity confinement. Instead, the mirror vibrations are induced by laser fields in a cold atomic-laser interactive setup \cite{PhysRevLett.107.223001, PhysRevA.82.021803, PhysRevA.93.023816}. In the proposed scheme, the atoms interact with two continuous laser fields: a pump and control fields. The control field is coupled with the mirror vibration before entering the atomic system (see Fig. \ref{heat_fig}). This coupling results in a phase modulation of the control field that produces sidebands in the control field by an amount of mirror frequency. The medium used in this scheme is EIT, known for its sensitivity to probe fields. Consequently, any perturbation in the system, such as the sidebands produced by the mirror vibration, can significantly influence the probe absorption spectrum. The integration of nanomechanical mirror vibrations with an EIT medium offers a promising approach for advancing quantum heat engines in optomechanical systems.

In this study, a theoretical model for three unconventional heat engines is presented. These engines utilize a four-level tripod-type atomic system that interacts with a nanomechanical mirror's simple harmonic oscillation and black body reservoirs. External forces acting on the mirror are neglected in this study, focusing solely on the interactions within the atomic system and mirror vibration. The control field sustains the constant mirror vibration within the trapped cold gas of atoms. The analysis focuses on studying the steady-state responses of all density matrix elements, which vibrate periodically due to the nanomechanical mirror's oscillation. The density matrix elements are expressed as Fourier series expansions, truncated up to the second order. Factors like mirror vibration, pump and control field strengths, and control field frequency are examined to understand their influence on the emitted output light, providing insights for performance optimization. Examining the output probe field variation amidst changing mirror vibrations reveals insights into their effects on characteristics. When considering the impact, it's noted that the three-level heat engine exhibits its highest brightness without mirror vibration. In contrast, the composite heat engine follows with the second-highest brightness output, which marginally rises with mirror vibration. Conversely, the three-level heat engine with mirror vibration experiences the lowest brightness, diminishing further with increasing mirror frequency. By adjusting the pump and control detuning/strength for different mirror frequencies, the engines are compared, and their effects are discussed. Comparing the proposed model of the three heat engines with the ideal system involves exploring entropy balance and investigating the limits imposed by thermodynamic laws on the temperature and entropy of emitted radiation. The model noted that mirror vibration results in a departure of the proposed heat engines from their ideal heat engine characteristics.

The article is structured as follows: The next section (\ref{theory}) discusses the details of the setup and the strategy of the model, followed by a description of the mathematical prerequisites of the theoretical model used for the present work in the subsequent section (\ref{tool}). The steady-state response of the system with a constant nanomechanical mirror is then analyzed in Section (\ref{result}). Finally, the study concludes in Section (\ref{con}).

\section{Setup and strategy}\label{theory}
\begin{figure*}[!th]
\begin{center}
\begin{tikzpicture}[
      scale=.3,
      level/.style={thick},
      virtual/.style={thick,densely dashed},
      trans/.style={thick,<->,shorten >=2pt,shorten <=2pt,>=stealth},
      photon/.style={thick,->,shorten >=0pt,decorate, decoration={snake}},
      classical/.style={thick,densely dashed,draw=blue}
       >=stealth',
  pos=.8,
  photon/.style={decorate,decoration={snake,post length=3mm}}
    ]   
    \draw [line width=0.5mm] [line width=0.8mm](24.5cm,-42.5em) -- (29.5cm,-42.5em) node[right] {$\vert 1\rangle $};
    \draw [line width=0.5mm] [line width=0.8mm](42cm,-35em) -- (47cm,-35em) node[right] {$\vert 3\rangle $};
    \draw [line width=0.5mm] [line width=0.8mm](33cm,-39em) -- (38cm,-39em) node[right] {$\vert 2\rangle $};
    \draw [line width=0.5mm] [line width=0.8mm](30cm,-15em) -- (40cm,-15em) node[right] {$\vert 4\rangle $};
 \draw [line width=2mm] [line width=2mm](19.7cm,20em) -- (19.7cm,2em) ;
 \draw [line width=0.5mm] [line width=0.5mm](19.8cm,-57em) -- (19.8cm,-61em) ;
\draw [line width=0.5mm] [line width=0.5mm](18cm,-53em) -- (18cm,-63em) ;

\node[] at (19.8cm,-52.5em) {$\textrm{z}_{0}$};
    \node[] at (66cm,-59.5em) {$\textrm{z}$};
      \node[] at (37cm,-17.5em) {$\Delta_{\textrm{c}}$};
       \node[] at (34cm,-17.5em) {$\Delta_{\textrm{pu}}$};
     \node[] at (26cm,-40.3em) {$\Delta_{\textrm{pr}}$};
      \node[] at (28cm,-29.5em) {$\textcolor{red}{\Omega_{\textrm{pr}}}$};
       \node[] at (29.3cm,-23em) {$\Gamma_{41}$};
        \node[] at (33cm,-25em) {$\Gamma_{42}$};
         \node[] at (39.9cm,-29em) {$\Gamma_{43}$};
      \node[] at (18cm,10em) {\textrm{M}};
    \node[] at (37cm,26.5em) {$\textcolor{magenta}{k_c, \omega_{c}}$};
     \node[] at (37cm,4em) {$\textcolor{blue}{k_{pu}, \omega_{pu}}$};
   \node[] at (42cm,-17.5em) {$+\omega_{\textrm{m}}$};
   \node[] at (42cm,-21.5em) {$-\omega_{\textrm{m}}$};
    \node[] at (60.5cm,-18em) {$L$};
    \node[] at (19.8cm,23em) {$\omega_{\textrm{m}}$};
    \node[] at (61cm,22.5em) {$\textcolor{red}{\textrm{T}_{41}}$};
     \node[] at (64cm,22.5em) {$\textcolor{red}{\textrm{T}_{42}}$};
    \node[] at (62cm,-8em) {$\textcolor{red}{\textrm{T}_{43}}$};

   \draw[trans] (27cm,-38em) -- (27cm,-43em) node[] {$ $};
   \draw[trans] (38cm,-20em) -- (38cm,-15em) node[] {$ $};
\fill[gray!50] (61,3) ellipse (7.5 and 2);
  \draw[][>=latex,thick,black](56,2.6) -- (50,-4)node[]{};
  \draw[][>=latex,thick,black](22,-4) -- (56,2.6)node[]{};
     \draw[][>=latex,thick,black](22,-4) -- (50,-4)node[]{};
      \draw[][>=latex,thick,black](22,-4) -- (22,-20)node[]{};
      \draw[][>=latex,thick,black](22,-20) -- (50,-20)node[]{};
       \draw[][>=latex,thick,black](50,-4) -- (50,-20)node[]{};
\filldraw [black] (57cm,5em) circle (8pt);
\filldraw [black] (58cm,5em) circle (8pt);
\filldraw [black] (59cm,5em) circle (8pt);
\filldraw [black] (56cm,7em) circle (8pt);
\filldraw [black] (63cm,6em) circle (8pt);
\filldraw [black] (61cm,7em) circle (8pt);
\filldraw [black] (60cm,5em) circle (8pt);
\filldraw [black] (64cm,8em) circle (8pt);
\filldraw [black] (59cm,8em) circle (8pt);
\filldraw [black] (57cm,8em) circle (8pt);
\filldraw [black] (59cm,8em) circle (8pt);
\filldraw [black] (62cm,8em) circle (8pt);
\filldraw [black] (66cm,9em) circle (8pt);
\filldraw [black] (65cm,11em) circle (8pt);
\filldraw [black] (57cm,11em) circle (8pt);
\filldraw [black] (59cm,11em) circle (8pt);
\filldraw [black] (62cm,11em) circle (8pt);
\draw[virtual][->,>=latex,thick,red](33,-5.2) -- (27,-13.2)node[]{};
\draw[virtual][->,>=latex,thick,black](34,-5.2) -- (34,-13.2)node[]{};
\draw[virtual][->,>=latex,thick,black](36,-5.2) -- (44,-12.3)node[]{};
\draw[virtual] (37cm,-17.5em) -- (40.5cm,-17.5em);
\draw[virtual] (36cm,-21.5em) -- (40.5cm,-21.5em);
\draw[virtual] (24.5cm,-38.5em) -- (29.5cm,-38.5em);
\draw[virtual] (34.5cm,-19.5em) -- (35.5cm,-19.5em);
\draw[virtual] (36cm,-19.5em) -- (40.5cm,-19.5em);
\draw[virtual] (19.5cm,10em) -- (19.1cm,19.8em);
\draw[virtual] (20cm,10em) -- (20.4cm,19.8em);
\path[draw=magenta,solid,line width=.5mm,fill=magenta,
preaction={stealth-stealth,very thick,draw,magenta,shorten >=-1mm}
] (44.5cm,-34.75em) -- (38.5cm,-20em) node[midway,right] {$\textcolor{magenta}{\Omega_{\textrm{c}}}$};
\path[draw=black,solid,line width=.5mm,fill=black,
preaction={stealth-stealth,very thick,draw,black,shorten >=-1mm}
] (66cm,-16em) -- (55cm,-16em) node[midway,right] {};
\path[draw=blue,solid,line width=0.5mm,fill=blue,
preaction={stealth-stealth,very thick,draw,blue,shorten >=-1mm}
] (35.5cm,-38.75em) -- (35.5cm,-20em) node[midway,right] {$\textcolor{blue}{\Omega_{\textrm{pu}}}$};
\path[draw=black,solid,line width=0.5mm,fill=black,
preaction={stealth-stealth,very thick,draw,black,shorten >=-1mm}
] (35.5cm,-20em) -- (35.5cm,-16em);
\path[draw=black,solid,line width=0.5mm,fill=black,
preaction={-latex,very thick,draw,black,shorten >=-1mm}
] (18cm,-59em) -- (65cm,-59em) node[] {};

\path[draw=magenta,solid,line width=0.5mm,fill=magenta,
preaction={latex-,very thick,draw,magenta,shorten >=-1mm}
] (50cm,28em) -- (70cm,38em) node[] {};
\path[draw=magenta,solid,line width=0.5mm,fill=magenta,
preaction={latex-,very thick,draw,magenta,shorten >=-1mm}
] (35cm,21em) -- (50cm,28em) node[] {};
\path[draw=magenta,solid,line width=0.5mm,fill=magenta,
preaction={latex-,very thick,draw,magenta,shorten >=-1mm}
] (20cm,15em) -- (35cm,21em) node[] {};
\path[draw=magenta,solid,line width=0.5mm,fill=magenta,
preaction={latex-,very thick,draw,magenta,shorten >=-1mm}
] (34.5cm,23em) -- (39.5cm,25em) ;

\path[draw=magenta,solid,line width=0.5mm,fill=magenta,
preaction={-latex,very thick,draw,magenta,shorten >=-1mm}
] (35cm,12.5em) -- (50cm,10em) node[] {};
\path[draw=magenta,solid,line width=0.5mm,fill=magenta,
preaction={-latex,very thick,draw,magenta,shorten >=-1mm}
] (20cm,15em) -- (35cm,12.5em) node[] {};

\path[draw=magenta,solid,line width=0.5mm,fill=magenta,
preaction={-latex,very thick,draw,magenta,shorten >=-1mm}
] (50cm,10em) -- (70cm,5em) node[] {};

\path[draw=blue,solid,line width=0.5mm,fill=blue,
preaction={latex-,very thick,draw,blue,shorten >=-1mm}
] (35cm,6.5em) -- (40cm,6em) node[] {};

\path[draw=blue,solid,line width=0.5mm,fill=blue,
preaction={-latex,very thick,draw,blue,shorten >=-1mm}
] (50cm,21em) -- (70cm,30em) node[] {};
\path[draw=blue,solid,line width=0.5mm,fill=blue,
preaction={-latex,very thick,draw,blue,shorten >=-1mm}
] (35cm,15em) -- (50cm,21em) node[] {};
\path[draw=blue,solid,line width=0.5mm,fill=blue,
preaction={-latex,very thick,draw,blue,shorten >=-1mm}
] (20cm,10em) -- (35cm,15em) node[] {};

\path[draw=blue,solid,line width=0.5mm,fill=blue,
preaction={latex-,very thick,draw,blue,shorten >=-1mm}
] (50cm,8em) -- (70cm,7em) node[] {};
\path[draw=blue,solid,line width=0.5mm,fill=blue,
preaction={latex-,very thick,draw,blue,shorten >=-1mm}
] (35cm,8.5em) -- (50cm,8em) node[] {};
\path[draw=blue,solid,line width=0.5mm,fill=blue,
preaction={latex-,very thick,draw,blue,shorten >=-1mm}
] (20cm,10em) -- (35cm,8.5em) node[] {};

\path[draw=red,solid,line width=2mm,fill=blue,
preaction={-latex,very thick,draw,red,shorten >=-3.5mm}
] (57cm,20em) -- (57cm,15em) node[] {};
\path[draw=red,solid,line width=2mm,fill=blue,
preaction={-latex,very thick,draw,red,shorten >=-3.5mm}
] (59cm,20em) -- (59cm,15em) node[] {};
\path[draw=red,solid,line width=2mm,fill=blue,
preaction={-latex,very thick,draw,red,shorten >=-3.5mm}
] (61cm,20em) -- (61cm,15em) node[] {};
\path[draw=red,solid,line width=2mm,fill=blue,
preaction={-latex,very thick,draw,red,shorten >=-3.5mm}
] (63cm,20em) -- (63cm,15em) node[] {};
\path[draw=red,solid,line width=2mm,fill=blue,
preaction={-latex,very thick,draw,red,shorten >=-3.5mm}
] (65cm,20em) -- (65cm,15em) node[] {};
\path[draw=red,solid,line width=2mm,fill=blue,
preaction={-latex,very thick,draw,red,shorten >=-4.3mm}
] (63cm,-5em) -- (63cm,0em) node[] {};
\path[draw=red,solid,line width=2mm,fill=blue,
preaction={-latex,very thick,draw,red,shorten >=-4.3mm}
] (61cm,-5em) -- (61cm,0em) node[] {};
\path[draw=red,solid,line width=2mm,fill=blue,
preaction={-latex,very thick,draw,red,shorten >=-4.3mm}
] (59cm,-5em) -- (59cm,0em) node[] {};
\path[draw=red,solid,line width=2mm,fill=blue,
preaction={-latex,very thick,draw,red,shorten >=-4.3mm}
] (57cm,-5em) -- (57cm,0em) node[] {};

\path[draw=red,solid,line width=2mm,fill=blue,
preaction={-latex,very thick,draw,red,shorten >=-4.3mm}
] (65cm,-5em) -- (65cm,0em) node[] {};
\end{tikzpicture}
\end{center}
\caption{ An ensemble of atoms (dots) is coupled by a mechanically oscillating mirror via pump (blue color line with wave number $k_{pu}$, frequency $\omega_{pu}$) and control (magenta color line with wave number $k_c$, frequency $\omega_{c}$) laser fields with two different optical paths. The control laser field is reflected from the mirror of mass $\textrm{M}$ before passing through the atomic system and the pump laser field, first passes through the atomic system, then hits the mirror, and finally exits from the system. The mirror vibrates with the frequency $\omega_{\textrm{m}}$ around its mean position $z_{0}$. The atoms are pumped by black body radiation at temperatures  $\textrm{T}_{41}$, $\textrm{T}_{42}$ and $\textrm{T}_{43}$. (Inset) The energy level diagram of the proposed tripod-type system, realized by pump and control laser fields of Rabi frequencies $\Omega_{\textrm{pu}}$ and $\Omega_{\textrm{c}}$, respectively. The pumping laser is in charge of connecting the transition $\ket{2}\leftrightarrow\ket{4}$, while the control laser drives the transition $\ket{3}\leftrightarrow\ket{4}$. The frequency detunings for the control and pump lasers are denoted by $\Delta_{\textrm{c}}$ and $\Delta_{\textrm{pu}}$, respectively. The black body radiation of temperatures $\textrm{T}_{41}$, $\textrm{T}_{42}$ and $\textrm{T}_{43}$ are pumping the transitions $\ket{1}\leftrightarrow\ket{4}$, $\ket{2}\leftrightarrow\ket{4}$ and $\ket{3}\leftrightarrow\ket{4}$, respectively. The Rabi frequency of the emitted probe field is $\Omega_{\textrm{pr}}$ with a frequency detuning denoted by $\Delta_{\textrm{pr}}$. The spontaneous decay rates from level $\ket{4}$ to levels $\ket{1}$, $\ket{2}$, and $\ket{3}$ are denoted by $\Gamma_{41}$, $\Gamma_{42}$, and $\Gamma_{43}$, respectively.} \label{heat_fig}
\end{figure*}
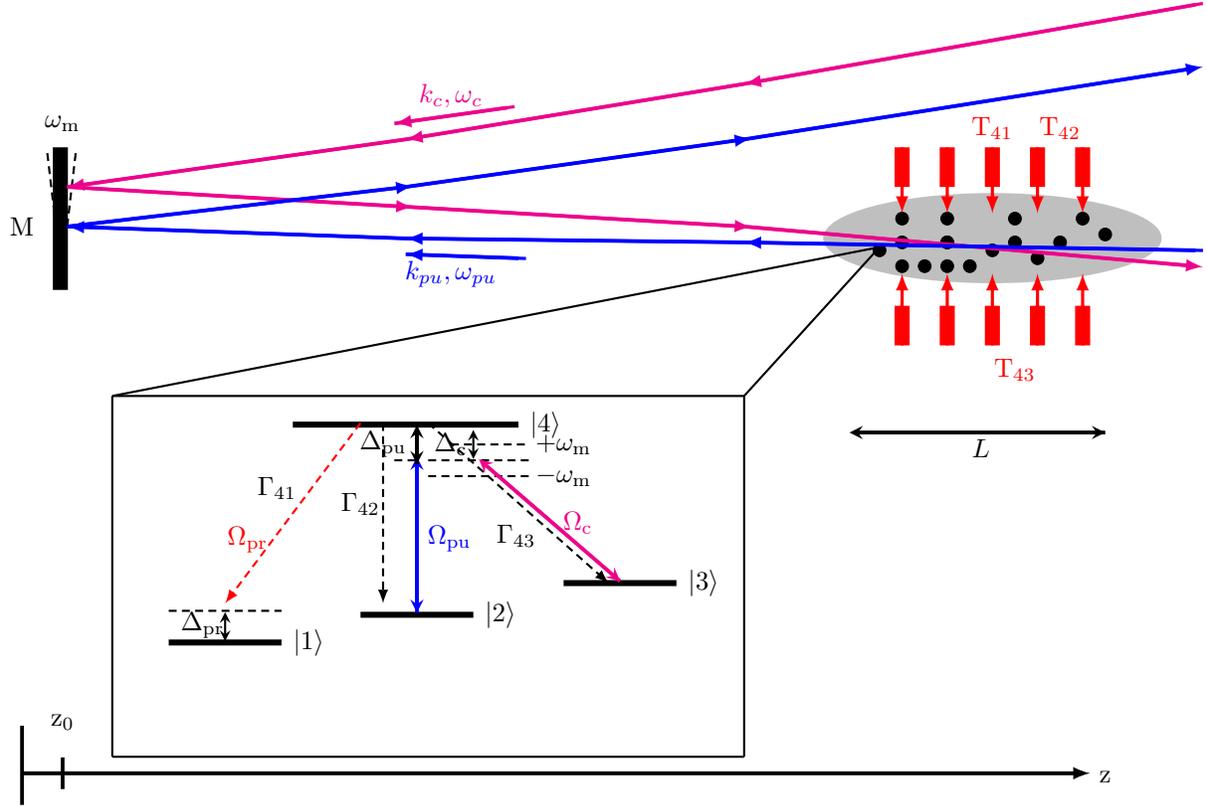

Fig. \ref{heat_fig} depicts the schematic of the proposed model system, which shows an ensemble of trapped, non-interacting cold atoms in contact with three blackbody reservoirs at temperatures of $\textrm{T}_{41}$, $\textrm{T}_{42}$ and $\textrm{T}_{43}$. The control laser field (wave number $k_{c}$, frequency $\omega_{c}$) reflected from the mirror of mass $\textrm{M}$ before passing through the atomic system; see magenta color line in Fig. \ref{heat_fig}. The pump laser field (wave number $k_{pu}$, frequency $\omega_{pu}$), first passes through the atomic system, then hits the mirror, and finally exits from the system; see the blue color line in Fig. \ref{heat_fig}. Hence, the effect of mirror vibration in the atomic system is carried out by the control laser field only. The mirror's centre of mass is positioned at $\textrm{z}=\textrm{z}_{0}$ and it oscillates with frequency $\omega_{\textrm{m}}$ around it's equilibrium position \textit{i.e.} at $\textrm{z}_{0}$. As shown in the inset of Fig. \ref{heat_fig}, the four atomic energy levels are considered, with one excited energy level $\ket{4}$ and three ground energy levels $\ket{1}$, $\ket{2}$, and $\ket{3}$ differentiated by the hyperfine levels of a definite fine level. The transitions $\ket{1}\rightarrow\ket{4}$, $\ket{2}\rightarrow\ket{4}$ and $\ket{3}\rightarrow\ket{4}$ are pumped using three blackbody reservoirs at temperatures $\textrm{T}_{41}$, $\textrm{T}_{42}$ and $\textrm{T}_{43}$, respectively. Photons are distributed in these three reservoirs at temperatures $\textrm{T}_{41}$, $\textrm{T}_{42}$ and $\textrm{T}_{43}$ according to Planck's distribution law \cite{planck1900theory}
\begin{eqnarray}\label{photon_dist}
\textrm{n}_{4i}=\frac{1}{\textrm{exp}^{\left(\frac{\hbar\omega_{4i}}{\textrm{K}_{\textrm{B}}\textrm{T}_{4i}}\right)}-1};~~ i=1, 2, 3
\end{eqnarray}
$\hbar=\frac{h}{2\pi}$, where $h$ and $\textrm{k}_{\textrm{B}}$ are the  Planck's constant and the Boltzmann constant, respectively. $\omega_{41}$ (or $\omega_{42}$ and $\omega_{43}$) is the characteristic frequency corresponding to the transition  $\ket{1}\rightarrow\ket{4}$ (or $\ket{2}\rightarrow\ket{4}$ and $\ket{3}\rightarrow\ket{4}$). The control field associates the transition $\ket{3}\rightarrow\ket{4}$ with the Rabi frequency $\Omega_{\textrm{c}}$ and the detuning $\Delta_{\textrm{c}}$. As per the proposed model, before entering the atomic system, the control field is reflected by the mirror. Thus the atoms will experience a time-dependent Rabi frequency of  $\Omega_{\textrm{c}}(t)$ which may be defined as
\begin{eqnarray}{\label{modulatedomega}}
    \Omega_{\textrm{c}}(t)= \Omega_{\textrm{c}}\textrm{exp}(ik_{c}\textrm{z}_{\textrm{m}}(t))\approx  \Omega_{\textrm{c}}[1+ik_{c}\textrm{z}_{\textrm{m}}(t)].
\end{eqnarray}
Here, the term $k_{c}\textrm{z}_{\textrm{m}}(t)$ is considered very small, \textit{i.e.} the control field wavelength ($\lambda_{\textrm{c}}=\frac{2\pi}{k_{c}}$) is large in comparison to the mirror displacement $\textrm{z}_{\textrm{m}}(t)$. The frequency of the control field $\omega_{c}$  is modified to $\omega_{c}\pm\omega_{\textrm{m}}$ for constant harmonic oscillation of the mirror, $\textrm{z}_{\textrm{m}}(t)=\textrm{z}_{0}\textrm{cos}(\omega_{\textrm{m}}t)$, as shown in the inset of Fig. \ref{heat_fig}. As a result, the control detuning is shifted by an amount $\pm\omega_{\textrm{m}}$, \textit{i.e.} $\Delta_{\textrm{c}}$ will be modified as $\Delta_{\textrm{c}}\rightarrow\Delta_{\textrm{c}}\pm\omega_{\textrm{m}}$. The pump field couples the transition $\ket{2}\rightarrow\ket{4}$ with the Rabi frequency $\Omega_{\textrm{pu}}$ and detuning $\Delta_{\textrm{pu}}$. 

The overall mechanism of the said model system is as follows in the presence of three blackbody reservoirs (at temperatures of $\textrm{T}_{41}$, $\textrm{T}_{42}$ and $\textrm{T}_{43}$) and the two laser fields (control and pump): The atoms are first pumped from level $\ket{1}$ to level $\ket{4}$ by the $\textrm{T}_{41}$ reservoir. As a result, photons are absorbed from the $\textrm{T}_{41}$ reservoir. The atoms in level $\ket{4}$ then spontaneously decay to level $\ket{3}$  (or $\ket{2}$), the emitted photons are absorbed by the $\textrm{T}_{43}$ (or $\textrm{T}_{42}$) reservoir. Further, the photons are absorbed from the control (or pump) laser field corresponding to the transition $\ket{3}\rightarrow\ket{4}$ (or $\ket{2}\rightarrow\ket{4}$), and therefore atoms from level $\ket{4}$ spontaneously emit photons at Rabi frequency $\Omega_{\textrm{pr}}$ for the $\ket{4}\rightarrow\ket{1}$ transition with detuning $\Delta_{\textrm{pr}}$. Here, the emitted photon is considered to have a single transverse mode, ignoring any scattering or reflection to other modes. The whole mechanism is surveyed by observing these emitted photons. It is worth noticeable that the control (or pump) laser field accounts for only a fraction $\frac{\omega_{43}}{\omega_{41}}$ (or $\frac{\omega_{42}}{\omega_{41}}$) of the power generated at $\omega_{41}$ \cite{4052124}. The main ingredient of the proposed model system is that the two applied laser fields (control and pump) along with the emitted probe field are operated under the regime of EIT. At the Raman resonance condition $\Delta_{\textrm{pr}}-\Delta_{\textrm{c}}\pm \omega_{\textrm{m}}=0$, the emitted photons are very sensitive to both the applied fields as well as to the mirror vibrations. 

In this scenario, three heat engine models are designed: $\textrm{HE}_{\textrm{pu}}$, $\textrm{HE}_{\textrm{c}}$, and $\textrm{HE}_{\textrm{pu,c}}$, each with specific characteristics and working principles.

    $(i)~\textrm{HE}_{\textrm{pu}}$:
    $\textrm{HE}_{\textrm{pu}}$ consists of two temperature reservoirs: $\textrm{T}_{41}$ and $\textrm{T}_{42}$. The heat engine cycle follows the nonlinear processes: $\ket{1} \xrightarrow{\textrm{T}_{41}} \ket{4} \xrightarrow{\textrm{T}_{42}} \ket{2} \xrightarrow{\Omega_{\textrm{pu}}} \ket{4} \xrightarrow{\Omega_{\textrm{pr}}} \ket{1}$.
    In this engine, the atoms produce low-grade work at the output photon mode $\Omega_{\textrm{pr}}$, while they take photons from the $\textrm{T}_{41}$ reservoir and release some into the $\textrm{T}_{42}$ reservoir.

    $(ii)~\textrm{HE}_{\textrm{c}}$:
    $\textrm{HE}_{\textrm{c}}$ is composed of two temperature reservoirs: $\textrm{T}_{41}$ and $\textrm{T}_{43}$, and involves mirror vibration. The heat engine cycle is supported by the nonlinear processes: $\ket{1} \xrightarrow{\textrm{T}_{41}} \ket{4} \xrightarrow{\textrm{T}_{43}} \ket{3} \xrightarrow{\Omega_{\textrm{c}}} \ket{4} \xrightarrow{\Omega_{\textrm{pr}}} \ket{1}$.
    In this case, the atoms generate low-grade work at the output photon mode $\Omega_{\textrm{pr}}$, taking photons from the $\textrm{T}_{41}$ reservoir and releasing some into the $\textrm{T}_{43}$ reservoir.

    $(iii)~\textrm{HE}_{\textrm{pu,c}}$:
    $\textrm{HE}_{\textrm{pu,c}}$ combines the features of both $\textrm{HE}_{\textrm{pu}}$ and $\textrm{HE}_{\textrm{c}}$ engines. It involves three temperature reservoirs: $\textrm{T}_{41}$, $\textrm{T}_{42}$, and $\textrm{T}_{43}$, along with mirror vibration. The heat engine cycle is supported by the nonlinear processes: $\ket{1} \xrightarrow{\textrm{T}_{41}~\textrm{and}~\textrm{T}_{41}} \ket{4} \xrightarrow{\textrm{T}_{43}~\textrm{and}~\textrm{T}_{42}} \ket{3}~\textrm{and}~\ket{2} \xrightarrow{\Omega_{\textrm{c}}~\textrm{and}~\Omega_{\textrm{pu}}} \ket{4} \xrightarrow{\Omega_{\textrm{pr}}~\textrm{and}~\Omega_{\textrm{pr}}} \ket{1}$.
    In this composite heat engine, the atoms produce low-grade work at the output photon mode $\Omega_{\textrm{pr}}$, taking photons from the $\textrm{T}_{41}$ reservoir and releasing some into both $\textrm{T}_{42}$ and $\textrm{T}_{43}$ reservoirs.

\section{Mathematical Prerequisites}\label{tool}

In this model of heat engines, a semi-classical approach is adopted, wherein the laser fields and the mirror vibrations are treated classically, while the atomic system is treated quantum mechanically. This choice is considered valid under two conditions: Firstly, the mirror is assumed to exhibit large-amplitude vibrations around its equilibrium position $\textrm{z}_{0}$. Secondly, the laser fields are assumed to be sufficiently intense to prevent quantum fluctuations. The model also neglects the travel time of the laser field between the mirror and atomic systems, which is much shorter than the mirror oscillation period $\textrm{T}_{m}=2\pi/\omega_{\textrm{m}}$. The mirror frequencies considered in the analysis are in the MHz range.

The oscillation of a classical mirror is explained using Newton's equation for a forced harmonic oscillator, which is described by the following equation:
\begin{eqnarray}{\label{Newton}}
    \textrm{M}\frac{d^2 \textrm{z}_{\textrm{m}}(t)}{d t^2}+\textrm{M}\omega_{\textrm{m}}^{2}\textrm{z}_{\textrm{m}}(t)=\textrm{F}_{\textrm{ext}}(t)
\end{eqnarray}
where, the external radiation force generated by the probe (emitted), control, and pump laser fields is given by $\textrm{F}_{\textrm{ext}}(t)$
\begin{eqnarray}
    \textrm{F}_{\textrm{ext}}(t)=\frac{2[\textrm{W}_{\textrm{pr}}(t)+\textrm{W}_{\textrm{c}}+\textrm{W}_{\textrm{pu}}]}{c}.
\end{eqnarray}

Because the emitted probe field generates within the atomic system before reflecting off the mirror, its power $\textrm{W}_{\textrm{pr}}(t)$ is time-dependent. The presence of both control and pump fields affects the transmission properties of emitted photons on the transition  $\ket{4}\rightarrow\ket{1}$. The transmitted power of the probe field through the atomic medium of length $L$ can be written as \cite{PhysRevA.93.023816}, 
\begin{eqnarray}{\label{mod_probe}}
     \textrm{W}_{\textrm{pr}}(t)=\textrm{W}_{\textrm{pr}}\textrm{exp}(-k_{pr}L\textrm{Im}[\rho_{41}^{em}])\approx\textrm{W}_{\textrm{pr}}(-k_{pr}L\textrm{Im}[\rho_{41}^{em}]))
\end{eqnarray}
where the $\rho_{jk}$ is the density matrix elements corresponding to the states $\ket{j}$ and $\ket{k}$; $k_{pr}$ is the wave no. of the emitted probe field. Since the probe field is generated within the medium, we only considered the modulated part in the last step of equation \eqref{mod_probe}. The power of the control and pump fields,  \textit{i.e.}, $\textrm{W}_{\textrm{c}}$ and $\textrm{W}_{\textrm{pu}}$, on the other hand, is constant.

The density matrix formalism is used to describe the dynamical response of the proposed model to laser fields and reservoirs, as well as the mirror. Assuming the dipole and rotating wave approximations, we can express the total Hamiltonian operator in the corotating frame for the $\textrm{HE}_{\textrm{pu,c}}$ engine as follows \cite{furman2016electromagnetically, laskar2021analysis}:
   \begin{eqnarray}
   \begin{aligned}
         \mathcal{H}=\hbar\lbrace (\Delta_{\textrm{pr}}-\Delta_{\textrm{pu}})\sigma_{22}+(\Delta_{\textrm{pr}}-\Delta_{\textrm{c}}\pm \omega_{\textrm{m}})\sigma_{33}+\Delta_{\textrm{pr}}\sigma_{44}\rbrace -\frac{\hbar}{2}\left[\Omega_{\textrm{pr}}\sigma_{14}+\Omega_{\textrm{pu}}\sigma_{24}+\Omega_{\textrm{c}}\sigma_{34}+h.c\right]
   \end{aligned}
     \end{eqnarray}
The transition operator $\sigma_{jk}=\vert j\rangle\langle k\vert$ denotes the transition between different states ($j\neq k$) and acts as a projection operator when $j=k$. The symbol $\hbar$  signifies the reduced Planck constant, while $\Delta_{\textrm{pr}}$ ($\Delta_{\textrm{c}}$ and $\Delta_{\textrm{pu}}$) stands for the frequency detuning related to the probe (control and pump) field. The $\Omega_{\textrm{pr}}$ ($\Omega_{\textrm{c}}$ and $\Omega_{\textrm{pu}}$) represents the Rabi frequency for probe (control and pump) field. $\omega_{m}$ represents the frequency of the mirror, and \textit{h.c} signifies the hermitian conjugate of the preceding terms enclosed within the third bracket. The Hamiltonian for the $\textrm{HE}_{\textrm{c}}$ and $\textrm{HE}_{\textrm{pu}}$ engines can be expressed as follows:
\begin{eqnarray}
\begin{aligned}
  \mathcal{H}&=\hbar\lbrace (\Delta_{\textrm{pr}}-\Delta_{\textrm{c}}\pm \omega_{\textrm{m}})\sigma_{33}+\Delta_{\textrm{pr}}\sigma_{44}\rbrace -\frac{\hbar}{2}\left[\Omega_{\textrm{pr}}\sigma_{14}+\Omega_{\textrm{c}}\sigma_{34}+h.c\right]~~\textrm{for}~ \textrm{HE}_{\textrm{c}}~\textrm{engine}
  \end{aligned}
  \end{eqnarray}
  and 
  \begin{eqnarray}
\begin{aligned}
   \mathcal{H}&=\hbar\lbrace (\Delta_{\textrm{pr}}-\Delta_{\textrm{pu}})\sigma_{22}+\Delta_{\textrm{pr}}\sigma_{44}\rbrace -\frac{\hbar}{2}\left[\Omega_{\textrm{pr}}\sigma_{14}+\Omega_{\textrm{pu}}\sigma_{24}+h.c\right]~~\textrm{for}~ \textrm{HE}_{\textrm{pu}}~\textrm{engine}
\end{aligned}
  \end{eqnarray}
  The detailed calculations leading to the expression of $\mathcal{H}$ are provided in Appendix I.  The density matrix elements are evolved according to the Lindblad master equation \textit{i.e.} 
\begin{eqnarray}\label{OBE}
\begin{aligned}
\partial_{t}\rho =\frac{i}{\hbar}[\rho,\mathcal{H}]+\mathcal{L}_{1}(\rho)+\mathcal{L}_{2}(\rho)
\end{aligned}
\end{eqnarray}
The terms $\mathcal{L}_{1}(\rho)$ and $\mathcal{L}_{2}(\rho)$ represent Lindblad superoperators, capturing dissipative effects due to spontaneous decay rates ($\Gamma_{4i}$) and interactions with temperature reservoirs ($\textrm{T}_{4i}$), respectively. The respective $\mathcal{L}_{1}(\rho)$ \cite{laskar2021analysis} and $\mathcal{L}_{2}(\rho)$ \cite{PhysRevLett.98.240601} are defined as
\begin{eqnarray}\label{decoherence}
\mathcal{L}_{1}(\rho)=\sum_{i}\frac{\Gamma_{4i}}{2}\left[2\sigma_{4i}\rho\sigma_{4i}-\sigma_{4i}\sigma_{4i}\rho-\rho\sigma_{4i}\sigma_{4i} \right].
\end{eqnarray}
and 
 \begin{eqnarray}\label{decoherence2}
\mathcal{L}_{2}(\rho)=\Gamma_{4i}(n_{4i}+1)\left(\left[\sigma_{4i}\rho,\sigma_{4i}^{\dagger}\right]-\left[\sigma_{4i}, \rho\sigma_{4i}^{\dagger}\right] \right)+\Gamma_{4i}n_{4i}\left(\left[\sigma_{4i}^{\dagger}\rho,\sigma_{4i}\right]-\left[\sigma_{4i}^{\dagger}, \rho\sigma_{4i}\right] \right)
\end{eqnarray}
$\Gamma_{4i}$, $\sigma_{4i}$, and $n_{4i}$ represent the spontaneous decay rates from level $\ket{4}$ to level $\ket{i}$, the transition operator from level $\ket{4}$ to level $\ket{i}$, and the photon number distribution in reservoir $\textrm{T}_{4i}$. The stimulated absorption or pumping rate $\textrm{R}_{i4}$ ($i=1, 2, 3$) from level $\ket{i}$ to level $\ket{4}$ due to the reservoirs $\textrm{T}_{4i}$ is expressed in terms of the corresponding spontaneous decay rate as \cite{PhysRevA.94.053859}, 
\begin{eqnarray}
\begin{aligned}
\textrm{R}_{i4}=\Gamma_{4i}n_{4i};~ i=1, 2, 3
\end{aligned}
\end{eqnarray}
   The dephasing rate ($\gamma_{jk}$) for the transitions $\ket{j}\rightarrow\ket{k}$ due to the spontaneous decay rate ($\Gamma_{4i}$) and temperature reservoir ($\textrm{T}_{4i}$) is obtained from the Lindblad superoperators $\mathcal{L}_{1}(\rho)$ [in equation \eqref{decoherence}] and $\mathcal{L}_{2}(\rho)$ [in equation \eqref{decoherence2}], respectively.
The dephasing rates for each of the transitions for composite $\textrm{HE}_{\textrm{pu,c}}$ engine can be written as \cite{PhysRevA.94.053859},
\begin{eqnarray}\label{dephasing}
\begin{aligned}
\gamma_{41}&=\Gamma_{41}+\Gamma_{42}+\Gamma_{43}+2 \textrm{R}_{14}+ \textrm{R}_{34}+ \textrm{R}_{24}\\ 
\gamma_{43}&=\Gamma_{41}+\Gamma_{42}+\Gamma_{43}+ \textrm{R}_{14}+2 \textrm{R}_{34}+\textrm{R}_{24}\\
\gamma_{42}&=\Gamma_{41}+\Gamma_{42}+\Gamma_{43}+ \textrm{R}_{14}+ \textrm{R}_{34}+2\textrm{R}_{24}\\
\gamma_{31}&= \textrm{R}_{14}+ \textrm{R}_{34}\\
\gamma_{21}&= \textrm{R}_{14}+\textrm{R}_{24}\\ 
\gamma_{32}&= \textrm{R}_{34}+\textrm{R}_{24}.
\end{aligned}
\end{eqnarray}
 In the first line, the dephasing rate $\gamma_{41}$ is comprised of the first three terms derived from $\mathcal{L}_{1}(\rho)$ and the last three terms obtained from $\mathcal{L}_{2}(\rho)$, \textit{i.e.}
 \begin{eqnarray}
\gamma_{41}&=\underbrace{\Gamma_{41}+\Gamma_{42}+\Gamma_{43}}_{\textrm{obtained from $\mathcal{L}_{1}(\rho)$}}+\underbrace{2 \textrm{R}_{14}+ \textrm{R}_{34}+ \textrm{R}_{24}}_{\textrm{obtained from $\mathcal{L}_{2}(\rho)$}}
 \end{eqnarray}
The specifics regarding dephasing rates for the $\textrm{HE}_{\textrm{pu}}$ and $\textrm{HE}_{\textrm{c}}$ engines are outlined in Appendix II.

In the present article, the driving force $\textrm{F}_{\textrm{ext}}(t)$ in Newton's equation [equation \eqref{Newton}] has been ignored. The oscillations of mirror, therefore, give rise constant side-band effects in control field \textit{i.e.} $ \Omega_{\textrm{c}}(t)\approx  \Omega_{\textrm{c}}[1+k_{c}\textrm{z}_{0}(\frac{\textrm{exp}(-i\omega_{\textrm{m}}t)+\textrm{exp}(i\omega_{\textrm{m}}t)}{2})]$ with side-band strength $k_{c}\textrm{z}_{0}$. This constant sideband in the control field prevents the atomic system from reaching a steady state condition. As a result, the elements of the density matrix are better represented by a periodic function with a periodicity of $\frac{2\pi}{\omega_{\textrm{m}}}$. Therefore, the density matrix elements can be expanded in a Fourier series \cite{PhysRevLett.96.184501} as,
\begin{eqnarray}\label{series}
    \Tilde{\rho}_{jk}=\sum_{l=-\infty}^{\infty}\Tilde{\rho}_{jk, l}\textrm{exp}(-il\omega_{\textrm{m}}t)
\end{eqnarray}
After a sufficient time, specifically when the transient phase has elapsed concerning the lifetime of the excited level $\ket{4}$ \textit{i.e.} $t\Gamma_{4i}\gg 1$ [i $\equiv$ 1,2,3], the amplitudes of Fourier series \textit{i.e.} $\Tilde{\rho}_{jk, l}$ in the steady state satisfies,
\begin{eqnarray}\label{steady}
\Dot{\Tilde{\rho}}_{jk, l}=0.
\end{eqnarray}

By solving the above equations, the constant side-bands generate an infinite number of coupled equations for $\Tilde{\rho}_{jk, l}$. By ignoring all $\Tilde{\rho}_{jk, l}$ terms for $\vert l\vert>1$, the constant density matrix elements $\Tilde{\rho}_{jk, 0}$ (the usual solution for $k_{c}\textrm{z}_{0}=0$) and the first harmonics in mirror frequency $\Tilde{\rho}_{jk, \pm}$ are retained in the Fourier expansion [Eq. \eqref{series}] as,
\begin{eqnarray}\label{density}
    \Tilde{\rho}_{jk}=\Tilde{\rho}_{jk, 0}+\Tilde{\rho}_{jk, +}\textrm{exp}(-i\omega_{\textrm{m}}t)+\Tilde{\rho}_{jk, -}\textrm{exp}(i\omega_{\textrm{m}}t).
\end{eqnarray}
The imaginary part of the probe coherence term (both absorption and emission) is modulated by these Fourier amplitudes ($\Tilde{\rho}_{jk, \pm}$). The probe emission term is defined as
\begin{eqnarray}
      \textrm{Im}[\Tilde{\rho}_{14}^{em}] =\textrm{Im}[\Tilde{\rho}_{14, 0}^{em}]+\delta\Tilde{\rho}_{14, 0}^{em}\textrm{cos}(\omega_{\textrm{m}}t+\alpha).
\end{eqnarray}
$\textrm{Im}[\Tilde{\rho}_{14, 0}^{em}]$ and $\alpha$ are the probe emission part without side-band effects and the phase difference between probe emission modulation and mirror motion, respectively. The amplitude of probe emission modulation is represented by $\delta\Tilde{\rho}_{14, 0}^{em}$. Using Eqs. \eqref{OBE} and \eqref{density}, the amplitudes of Fourier series are obtained for $\textrm{HE}_{\textrm{pu,c}}$ engine 
\begin{eqnarray}\label{mainsolution}
    \Tilde{\rho}_{14, \pm}=\frac{i\Omega_{\textrm{pr}}}{2G}(\rho_{11}-\rho_{44})+\frac{\Omega_{\textrm{pr}}\Omega_{\textrm{c}}\frac{k_{0}z_{0}}{2}\Tilde{\rho}_{43, \pm}}{4(\frac{1}{2}\gamma_{31}
-i(\Delta_{\textrm{pr}}-\Delta_{\textrm{c}})\mp i\omega_{\textrm{m}})G}
\end{eqnarray}
and 
\begin{eqnarray}\label{mainsolution2}
    \Tilde{\rho}_{43, \pm}=\frac{i(\frac{k_{0}\textrm{z}_{0}}{2F})\Omega_{\textrm{c}}}{2F}(\rho_{44}-\rho_{33})+\frac{\Omega_{\textrm{pr}}\Omega_{\textrm{c}}\frac{k_{0}z_{0}}{2}\Tilde{\rho}_{14, \pm}}{4(\frac{1}{2}\gamma_{31}
-i(\Delta_{\textrm{pr}}-\Delta_{\textrm{c}})\mp i\omega_{\textrm{m}})F}
\end{eqnarray}
 The $\Tilde{\rho}_{jk, l}$ retained up to the first order in $\Omega_{\textrm{pr}}$ and $\Omega_{\textrm{c}}$. The $G$, $F$ and populations terms are detailed in Appendix III. The terms $\alpha$ and $\delta\Tilde{\rho}_{14,0}^{em}$ are defined as
\begin{eqnarray}
\begin{aligned}
\alpha=&\textrm{Argument}[\Tilde{\rho}_{14, +}^{\textrm{em}}-\Tilde{\rho}_{14, -}^{*\textrm{em}}]+\frac{\pi}{2}\\
\delta\Tilde{\rho}_{14,0}^{em}&=\textrm{Modulas}[\Tilde{\rho}_{14,+}^{\textrm{em}}-\Tilde{\rho}_{14,-}^{*\textrm{em}}]
\end{aligned}
\end{eqnarray}
 
The spectral brightness $\textrm{B}(\omega_{\textrm{pr}},\textrm{z})$ of single mode emitted photons in the $\textrm{z}$ direction follows the equation \cite{PhysRevA.94.053859, 1984oup..book.....M}
\begin{eqnarray}\label{eqb}
    \frac{\textrm{d}\textrm{B}(\omega_{\textrm{pr}},\textrm{z})}{\textrm{dz}}+[\sigma_{\textrm{abs}}-\sigma_{\textrm{em}}]\textrm{B}(\omega_{\textrm{pr}},\textrm{z})=\sigma_{\textrm{em}}.
\end{eqnarray}
 The absorption and emission coefficients are respectively defined as  
\begin{eqnarray}\label{crosssection}
\begin{aligned}
    \sigma_{i}=\textrm{Im}[\Tilde{\rho}_{14}^{i}];~~(i=\textrm{abs},\textrm{em})
    \end{aligned}
\end{eqnarray}
In the $\textrm{HE}_{\textrm{pu,c}}$ engine, $\sigma_{\textrm{abs}}$ depends on $\rho_{11}$, while $\sigma_{\textrm{em}}$ depends on $\rho_{22}$, $\rho_{33}$, and $\rho_{44}$. In the $\textrm{HE}_{\textrm{pu}}$ [or $\textrm{HE}_{\textrm{c}}$] engine, $\sigma_{\textrm{abs}}$ depends on $\rho_{11}$, while $\sigma_{\textrm{em}}$ depends on $\rho_{22}$ [ or $\rho_{33}$] and $\rho_{44}$.
 Solving Eq. \eqref{eqb} with boundary condition  $\textrm{B}(\omega_{\textrm{pr}},0)=0$, the brightness at a distance $\textrm{z}$ can be written as 
\begin{eqnarray}
     \textrm{B}(\omega_{\textrm{pr}},\textrm{z})=\textrm{B}(\omega_{\textrm{pr}},\infty)[1-e^{-\frac{\textrm{z}}{\textrm{z}_{0}}}].
\end{eqnarray}
where,
\begin{eqnarray}
     \textrm{B}(\omega_{\textrm{pr}},\infty)=\frac{\sigma_{\textrm{em}}}{(\sigma_{\textrm{abs}}-\sigma_{\textrm{em}})}.
\end{eqnarray}
Beyond the characteristic length \textit{i.e.} $\textrm{z}>\textrm{z}_{0}$, the brightness approaches its asymptotic limit $\textrm{B}(\omega_{\textrm{pr}},\infty)$ and this entity is considered as the brightness in the present article and denoted by $\textrm{B}(\Delta_{\textrm{pr}})$.

\section{ANALYSIS}\label{result}
In this section, the spectral normalized brightness of the emitted probe field is compared for the previously proposed heat engines, namely $\textrm{HE}_{\textrm{pu}}$, $\textrm{HE}_{\textrm{c}}$, and $\textrm{HE}_{\textrm{pu,c}}$. The effect of mirror vibration, pump, and control field strengths on the spectral normalized brightness at the line center ($\Delta_{\textrm{pr}}=0$) is investigated. Furthermore, the effect of control detuning on the spectral normalized brightness at the line center is explored for all the proposed heat engines. The study also examines the amplitude of probe emission modulated term and its phase difference with mirror motion. Finally, an entropy balance analysis is conducted to compare all suggested engines with the ideal engine. Moreover, the entropy flow rate per unit power of the emitted probe field is compared for all three suggested heat engines. In the analysis, the normalized brightness refers to the brightness [$\textrm{B}(\Delta_{\textrm{pr}})$] of the emitted probe field \textit{w.r.t.} the photon occupation number $\textrm{n}_{41}$ of the $\textrm{T}_{41}$ reservoir.

For experimental realization, I have implemented my model in the D$_{1}$-line of the $^{87}$Rb atomic system. The levels $\ket{1}$, $\ket{2}$ and $\ket{3}$ are assign to the $ 5S_{\frac{1}{2}}$ level with $F=1, m_{F}=0$ and $F=2, m_{F}=(-2,0)$, respectively. The level $\ket{4}$ is equivalent to $ 5P_{\frac{1}{2}}$ with $F=2, m_{F}=-1$ \cite{PhysRevA.89.021802}. The characteristic frequencies are  $\omega_{31}=6.831$ (2$\pi$. GHz), $\omega_{41}=4\times 10^{15}$ (2$\pi$. Hz) and $\omega_{43}=3\times 10^{15}$ (2$\pi$. Hz). Here I have assumed that $\omega_{43}=\omega_{42}$. The intensity of the laser field in terms of Rabi frequency $\Omega$ can be written as \cite{steck2001rubidium}, 
\begin{eqnarray}
    I=2I_{s}\left(\frac{2\Omega}{\Gamma}\right)^{2}
\end{eqnarray}
where, 
\begin{eqnarray}
    I_{s}=\frac{c\epsilon_{0}\Gamma^{2}\hbar^{2}}{4\vert \Vec{e}.\Vec{d}\vert^{2}}
\end{eqnarray}
$I_{s}$, $\Gamma$, $\Vec{e}$, and $\Vec{d}$  are saturation intensity, natural decay rate, polarization vector, and dipole moment, respectively. For D$_{1}$-line, the saturation intensity and natural decay rates are $I_{s}=4.484 mW/cm^{2} $and $\Gamma=\Gamma_{41}=\Gamma_{42}=\Gamma_{43}=\frac{1}{3}\times5.7$ (2$\pi$. MHz) \cite{steck2001rubidium}. The Rabi frequency corresponds to  $I_{s}=4.484 mW/cm^{2}$ is $\Omega=4$ (2$\pi$. MHz) \cite{steck2001rubidium}. The temperatures of the black body reservoirs are considered as, $\textrm{T}_{41}=\textrm{T}_{42}=\textrm{T}_{43}=$5000 K \cite{PhysRevA.103.062205, PhysRevLett.119.050602}. The dephasing rates are obtained from equation \eqref{dephasing} as follows: $\gamma_{21}=0.072$ (2$\pi$. MHz), $\gamma_{32}=0.118$ (2$\pi$. MHz), $\gamma_{31}=0.131$ (2$\pi$. MHz), $\gamma_{41}=17.360$ (2$\pi$. MHz), $\gamma_{42}=17.410$ (2$\pi$. MHz), $\gamma_{43}=17.410$ (2$\pi$. MHz). The three screening filters are working on pumping reservoirs, which are not shown in Fig.\ref{heat_fig}. Phase matching of these two lasers (control and pump) is not considered, and the directions are arbitrary.

\subsection{Normalized brightness of emitted probe field}

In this subsection, the spectral normalized brightness of the emitted probe field in the proposed model of heat engines is discussed. The engines are operating under the typical EIT conditions. At the end of this subsection, the investigation reveals which heat engine yields more output photons \textit{w.r.t.} the photon occupation number ($\textrm{n}_{41}$) of the $\textrm{T}_{41}$ reservoir. In all calculations involving normalized brightness, absorption, or emission coefficients, the constant density matrix elements $\Tilde{\rho}_{jk, 0}$ (representing the typical solution for $k_{c}\textrm{z}_{0}=0$) and the first harmonics of both positive and negative mirror frequencies, \textit{i.e.}, $\Tilde{\rho}_{jk, \pm}$, are considered in the Fourier expansion as mentioned in equation \eqref{density}. However, in the $\textrm{HE}_{\textrm{pu}}$ engine, only the constant density matrix element $\Tilde{\rho}_{jk, 0}$ is considered, as it is designed in the absence of mirror vibration.

\begin{figure}
\centering
\includegraphics[scale=0.55]{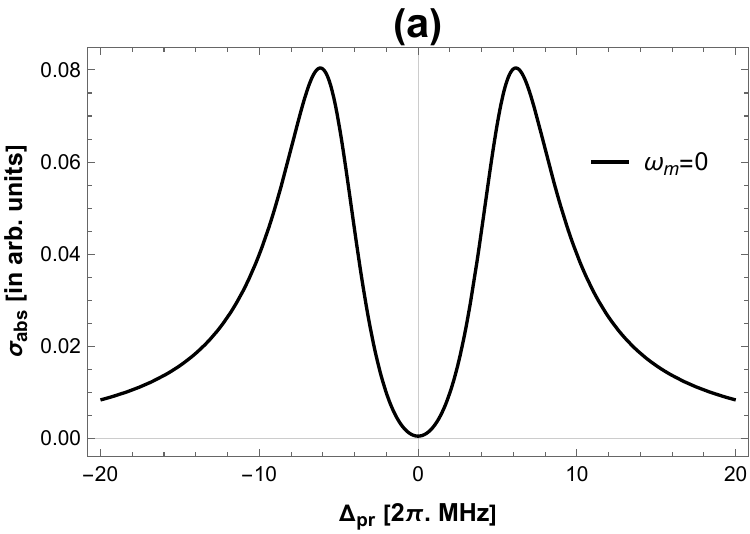}
\includegraphics[scale=0.55]{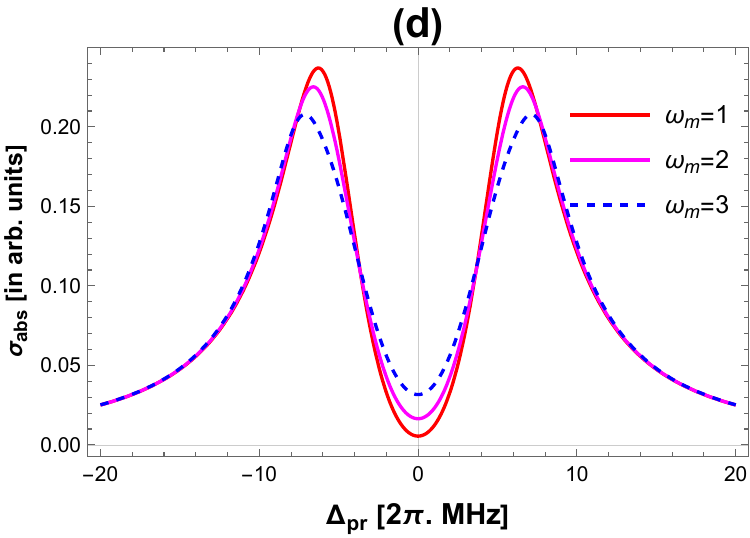}
 \includegraphics[scale=0.55]{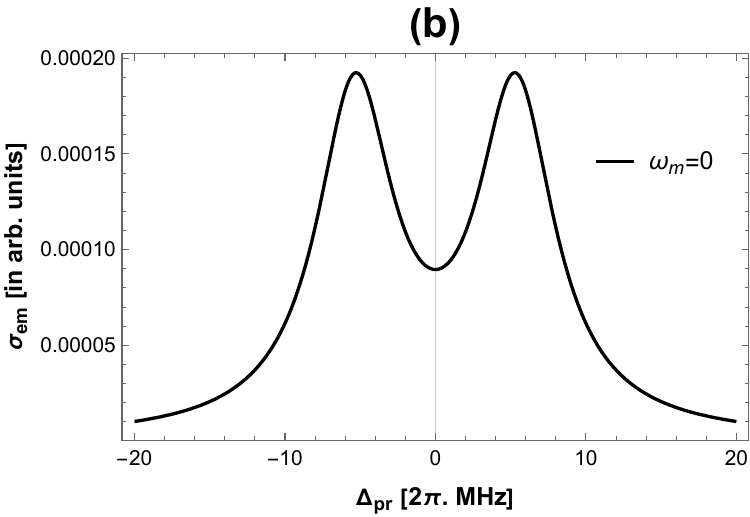}
 \includegraphics[scale=0.55]{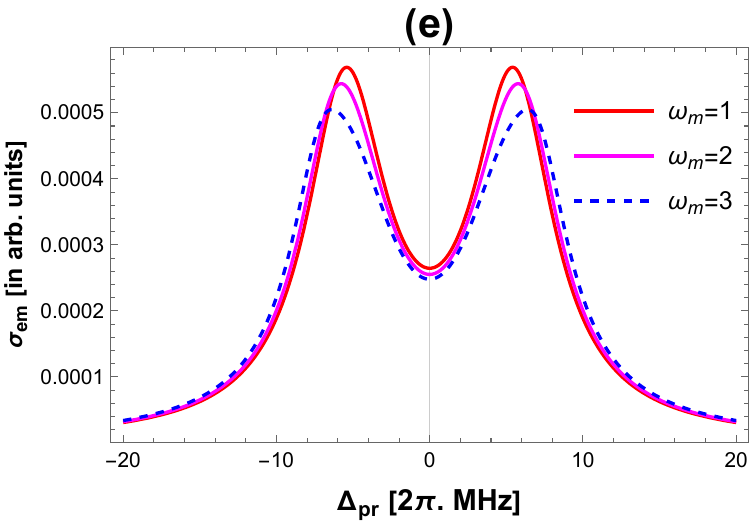}
 \includegraphics[scale=0.55]{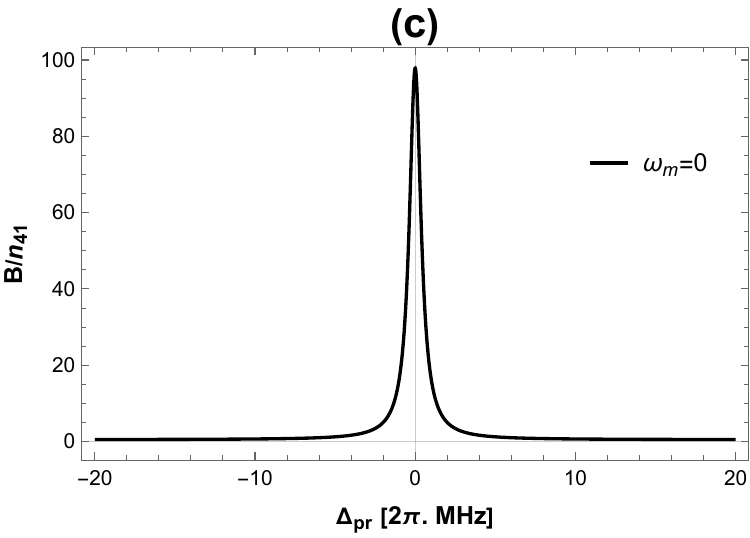}
\includegraphics[scale=0.55]{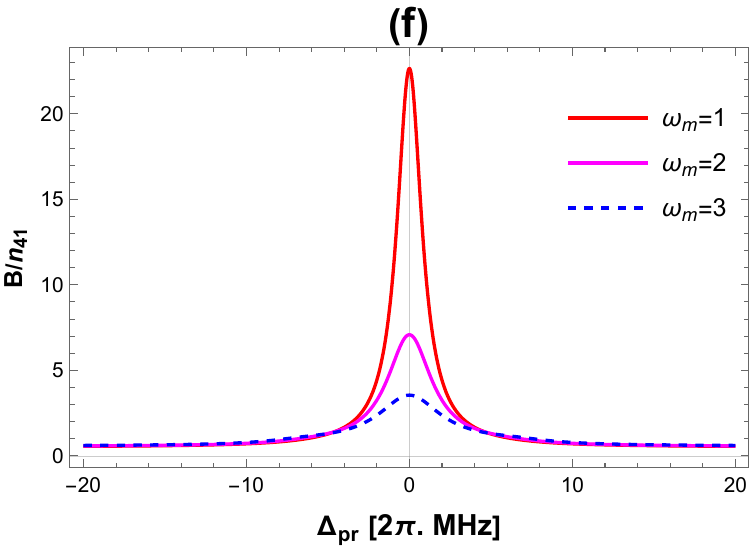}
\caption{The probe absorption coefficient $\sigma_{\textrm{abs}}$ [in plots (a) and (d)], emission coefficient $\sigma_{\textrm{em}}$ [in plots (b) and (e)] and normalized brightness $\frac{\textrm{B}(\Delta_{\textrm{pr}})}{\textrm{n}_{41}}$  [in plots (c) and (f)] are depicted as a function of probe detuning ($\Delta_{\textrm{pr}}$) for \textrm{HE}$_{\textrm{pu}}$ [in the left panel \textit{i.e.} (a), (b) and (c)] and \textrm{HE}$_{\textrm{c}}$ [in the right panel \textit{i.e.} (c), (e) and (f)] engines. The solid black color (in the left panel), solid red color (in the right panel), solid magenta color (in the right panel), and dashed blue color (in the right panel) curves are for  $\omega_{\textrm{m}}=0$,  $\omega_{\textrm{m}}=1$ (2$\pi$. MHz),  $\omega_{\textrm{m}}=2$ (2$\pi$. MHz), and  $\omega_{\textrm{m}}=3$ (2$\pi$. MHz), respectively. In the left panel, the parameters are $\Omega_{\textrm{pu}}=\gamma_{41}$, $\textrm{T}_{41}$=5000 K, $\textrm{T}_{42}$=5000 K, $\Delta_{\textrm{pu}}=0$, $\omega_{41}=4\times10^{15}$ (2$\pi$. Hz) and $\omega_{42}=3\times10^{15}$ (2$\pi$. Hz). In the right panel, the parameters are  $\Omega_{c}=\gamma_{41}$, $\textrm{T}_{43}$=5000 K, $\omega_{43}=3\times10^{15}$ (2$\pi$. Hz), $\Delta_{\textrm{c}}=0$, and $k_{0}z_{0}=0.01$.  \label{proberesponse}}
\end{figure}

In Fig. \ref{proberesponse}, Figs. \ref{proberesponse}(a) and \ref{proberesponse}(d) display the probe absorption coefficient $\sigma_{\textrm{abs}}$, Figs. \ref{proberesponse}(b) and \ref{proberesponse}(e) depict the emission coefficient $\sigma_{\textrm{em}}$, and Figs. \ref{proberesponse}(c) and \ref{proberesponse}(f) present the normalized spectral brightness $\frac{\textrm{B}(\Delta_{\textrm{pr}})}{\textrm{n}_{41}}$. The left panel illustrates the steady state response of the $\textrm{HE}_{\textrm{pu}}$ engine, while the right panel exhibits the steady state response of the $\textrm{HE}_{\textrm{c}}$ engine. Within these panels, the solid black curve (left panel), solid red curve (right panel), solid magenta curve (right panel), and dashed blue curve (right panel) represent $\omega_{\textrm{m}}$ values of 0, 1 (2$\pi$. MHz), 2 (2$\pi$. MHz), and 3 (2$\pi$. MHz) respectively.

Both the absorption coefficient ($\sigma_{\textrm{abs}}$) and the emission coefficient ($\sigma_{\textrm{em}}$) for the $\textrm{HE}_{\textrm{pu}}$ engine exhibit an EIT window around $\Delta_{\textrm{pr}}=0$ [see the Figs. \ref{proberesponse}(a) and \ref{proberesponse}(b)]. The magnitude of the absorption profile is 100 times greater than that of the emission profile, as depicted in Figs. \ref{proberesponse}(a) and \ref{proberesponse}(b). Therefore, the absorption profile plays a crucial role in changing the brightness of emitted light compared to the emission profiles. This discrepancy arises from the higher population in level $\rho_{11}$ (contributing to the absorption profile of $\textrm{HE}_{\textrm{pu}}$ engine) compared to the population in levels $\rho_{22}$ and $\rho_{44}$ (contributing to the emission profiles of $\textrm{HE}_{\textrm{pu}}$ engine). The brightness of light is determined by the balance between emission and absorption. In EIT conditions, where absorption nearly diminishes within the probe transition, this becomes the pivotal condition for significantly brighter emitted probe light. Consequently, brighter emitted light relative to photon no. (here $\textrm{n}_{41}$) of the respective temperature reservoirs (here $\textrm{T}_{41}$ reservoir) is achieved, as demonstrated in Haris' work \cite{PhysRevA.94.053859} and depicted in this study, as shown in Fig. \ref{proberesponse}(c). In the right panel, the $\textrm{HE}_{\textrm{c}}$ engine is under consideration, introducing the influence of mirror frequency. When the mirror frequency is $\omega_{m}=1$ (2$\pi$. MHz), the EIT window is at its maximum in the $\sigma_{\textrm{abs}}$ profile (solid red curve in \ref{proberesponse}(d)), while it appears at a minimum in the $\sigma_{\textrm{em}}$ profile (solid red curve in \ref{proberesponse}(e)). Thus, the $\textrm{HE}_{\textrm{c}}$ engine emits brighter light compared to the $\textrm{n}_{41}$ of the $\textrm{T}_{41}$ reservoir. This is illustrated in the solid red curve of Fig. \ref{proberesponse}(f), indicating that its EIT window is at a maximum, signifying nearly zero absorption. As the mirror frequency increases to $\omega_{m}=2$ (2$\pi$. MHz) and $\omega_{m}=3$ (2$\pi$. MHz), the EIT window decreases in the $\sigma_{\textrm{abs}}$ profile (solid magenta and dashed blue curves in \ref{proberesponse}(d)). Consequently, the normalized brightness of emitted probe light decreases with increasing mirror frequency, as evident from the solid magenta and dashed blue curves in Fig. \ref{proberesponse}(f). The normalized brightness reaches its maximum when there's no mirror frequency present (as observed in the curve in Fig. \ref{proberesponse}(c)). However, it decreases as the mirror frequency increases, as depicted by the curves in Fig. \ref{proberesponse}(f). This decrease occurs because the mirror frequency reduces the EIT window in the absorption profile. Moreover, as evident in Figs. \ref{proberesponse}(c) and \ref{proberesponse}(f), the width of the normalized brightness profiles expands as the mirror frequency increases. This expansion signifies that the frequency range of the emitted probe field, brighter than $\textrm{n}_{41}$, expands with the presence of mirror frequency or vibration. This contributes to the decrease in the normalized brightness profile with increasing mirror frequency.

\begin{figure}
\centering
\includegraphics[scale=0.55]{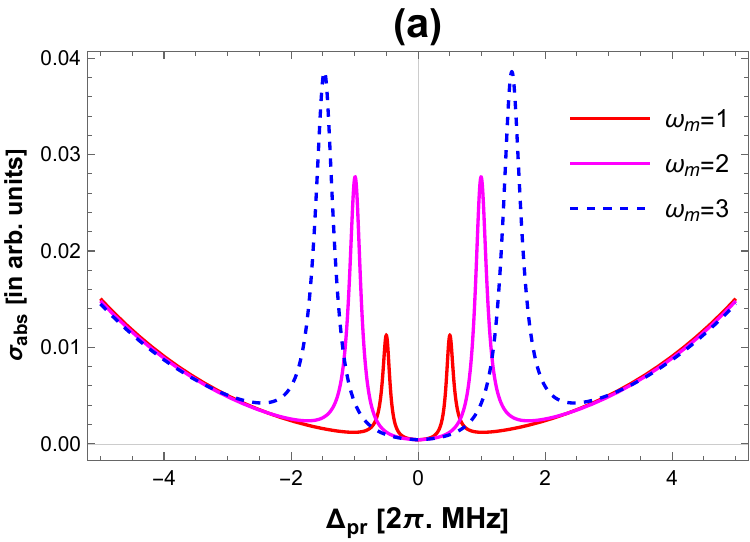}
\includegraphics[scale=0.57]{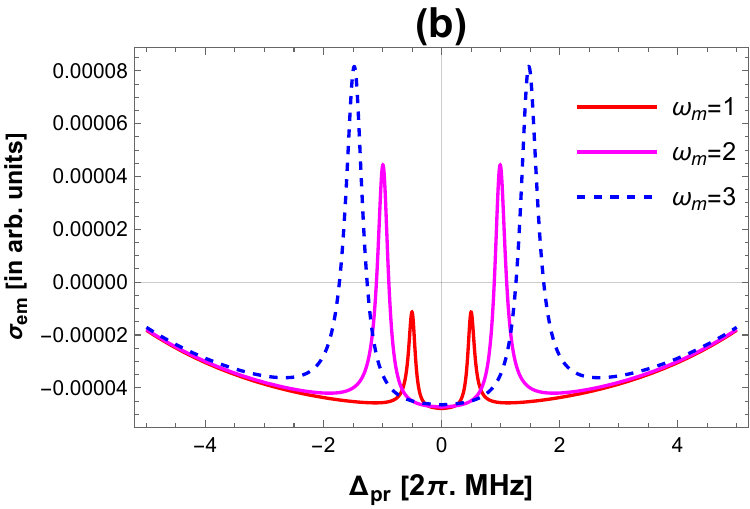}
\includegraphics[scale=0.55]{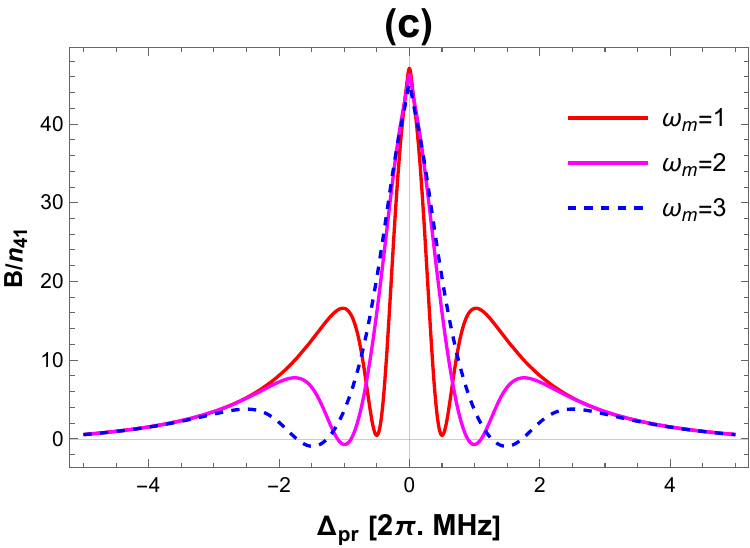}
\caption{ The probe absorption coefficient $\sigma_{\textrm{abs}}$ [in the plot(a)], emission coefficient $\sigma_{\textrm{em}}$ [in the plot(b)] and normalized brightness $\frac{\textrm{B}(\Delta_{\textrm{pr}})}{\textrm{n}_{41}}$ [in the plot (c)] are depicted as a function of probe detuning ($\Delta_{\textrm{pr}}$) for composite \textrm{HE}$_{\textrm{pu,c}}$ engine. The solid red color, solid magenta color, and dashed blue color curves are for    $\omega_{\textrm{m}}=1$ (2$\pi$. MHz),  $\omega_{\textrm{m}}=2$ (2$\pi$. MHz), and  $\omega_{\textrm{m}}=3$ (2$\pi$. MHz), respectively. All other parameters are  $\Omega_{\textrm{pu}}=\gamma_{41}$, $\Omega_{c}=\gamma_{41}$, $\textrm{T}_{41}=\textrm{T}_{42}=\textrm{T}_{43}$=5000 K, $\omega_{41}=4\times10^{15}$ (2$\pi$. Hz), $\omega_{42}=\omega_{43}=3\times10^{15}$ (2$\pi$. Hz), $\Delta_{\textrm{pu}}=0$, $\Delta_{\textrm{c}}=0$ and $k_{0}z_{0}=0.01$. \label{proberesponse_HAB}}
\end{figure}

In Fig.\ref{proberesponse_HAB}, the variations of the probe absorption coefficient $\sigma_{\textrm{abs}}$ (see Fig. \ref{proberesponse_HAB}(a)), emission coefficient $\sigma_{\textrm{em}}$ (see to Fig. \ref{proberesponse_HAB}(b)), and the normalized spectral brightness $\frac{\textrm{B}(\Delta_{\textrm{pr}})}{\textrm{n}_{41}}$ (depicted in Fig. \ref{proberesponse_HAB}(c)) are presented against the probe detuning $\Delta_{\textrm{pr}}$. These results are obtained using the composite $\textrm{HE}_{\textrm{pu,c}}$ engine. The solid red, solid magenta, and dashed blue curves correspond to $\omega_{\textrm{m}}$ values of 1 (2$\pi$. MHz), 2 (2$\pi$. MHz), and 3 (2$\pi$. MHz), respectively. In Figure \ref{proberesponse_HAB}(a), two absorption peaks have been observed. The separation between these peaks and their amplitudes increases as the mirror frequency rises. A similar pattern is evident in the emission profiles shown in Fig. \ref{proberesponse_HAB}(b). This occurs due to the presence of a cross-phase interaction between the control field and the emitted probe field.  In Figure \ref{proberesponse_HAB}(b), the emission profiles display negative values, signifying atom loss during the emission process. This outcome arises from accounting for all dephasing rates ($\gamma_{21}$, $\gamma_{32}$, and $\gamma_{31}$) between the three dipole forbidden levels $\ket{1}$, $\ket{2}$, and $\ket{3}$ within the $\textrm{HE}_{\textrm{pu,c}}$ engine. The increased absorption within the absorption profile due to the rising mirror frequency [see Fig.\ref{proberesponse_HAB}(a)] results in emitted light that is less bright [see Fig.\ref{proberesponse_HAB}(c)]. Additionally, the widening separation between the absorption profiles with increasing mirror frequency [see Fig.\ref{proberesponse_HAB}(a)] causes the brightness profiles to become wider with increasing mirror frequency [see Fig.\ref{proberesponse_HAB}(c)]. Therefore, in Figure \ref{proberesponse_HAB}(c), a marginal rise in amplitude is observed with the decreasing mirror frequency, and there's a noticeable expansion in the width of the normalized brightness profile as the mirror frequency rises. This widening of the brightness profile indicates a broader spectral region of the emitted light brighter than $\textrm{n}_{41}$. Therefore, the mirror frequency marginally improves the normalized brightness for the composite $\textrm{HE}_{\textrm{pu,c}}$ engine, whereas it leads to a reduction in normalized brightness for the $\textrm{HE}_{\textrm{c}}$ engine.

In summary, the heat engine $\textrm{HE}_{\textrm{pu}}$ exhibits the highest normalized brightness [see to Fig. \ref{proberesponse}(c)], while the composite heat engine $\textrm{HE}_{\textrm{pu,c}}$ displays the second-highest normalized brightness [see Fig. \ref{proberesponse_HAB}(c)]. This brightness slightly increases with mirror frequency. Conversely, in $\textrm{HE}_{\textrm{c}}$ engine, the normalized brightness reaches its minimum value [see Fig. \ref{proberesponse}(f)] and decreases with increasing mirror frequency.

\subsection{Effect of the pump, control field strengths, and control detuning on normalized brightness}

\begin{figure}
  \centering
  \includegraphics[width=0.43\linewidth]{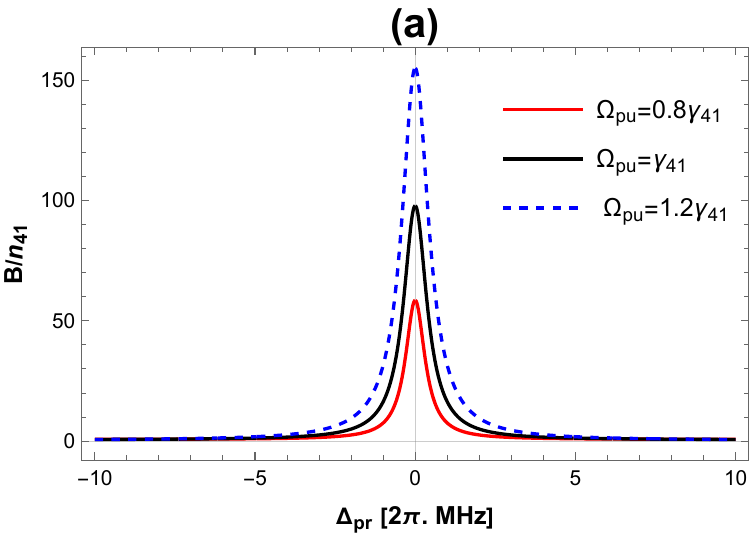}
 \includegraphics[width=0.43\linewidth]{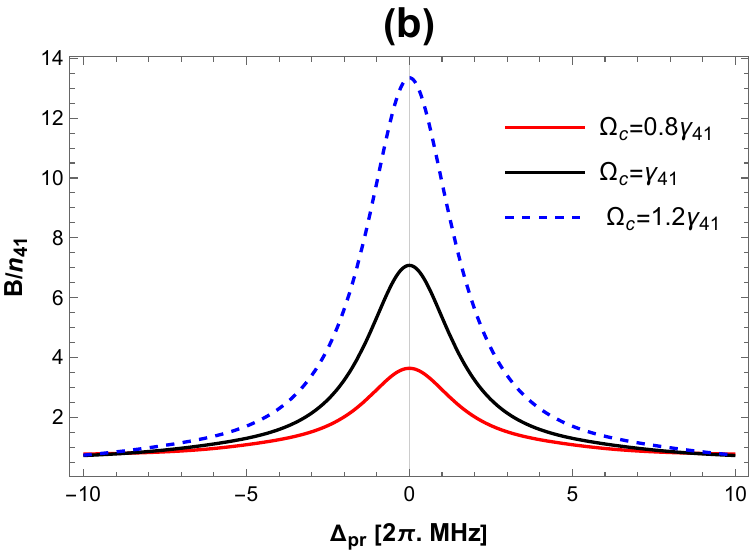}
   \includegraphics[width=0.43\linewidth]{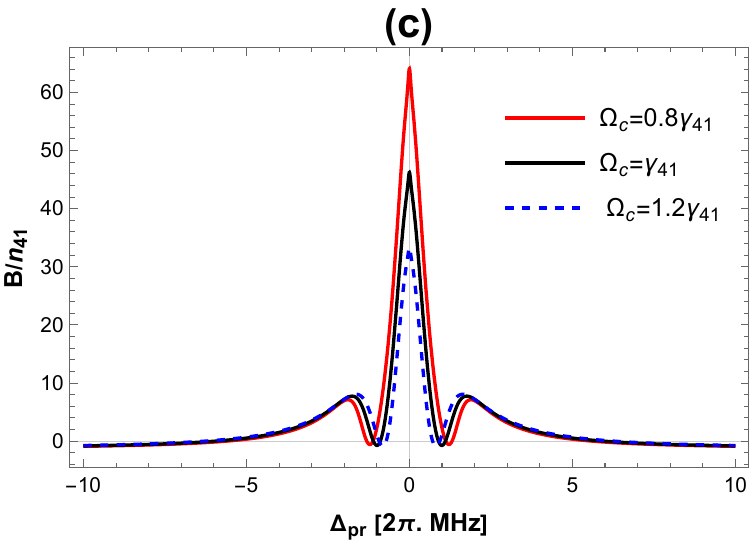}
  \caption{ The normalized brightness, denoted as $\frac{\textrm{B}(\Delta_{\textrm{pr}})}{\textrm{n}_{41}}$, is depicted against the probe detuning ($\Delta_{\textrm{pr}}$) for three different heat engines: (a) \textrm{HE}$_{\textrm{pu}}$, (b) \textrm{HE}$_{\textrm{c}}$, and (c) \textrm{HE}$_{\textrm{pu,c}}$. In plot (a), the curves are distinguished by solid red, solid black, and dashed blue colors, corresponding to $\Omega_{\textrm{pu}}$ values of $0.8\gamma_{41}$, $\gamma_{41}$, and $1.2\gamma_{41}$, respectively. Similarly, in plots (b) and (c), the curves are colored in solid red, solid black, and dashed blue, indicating $\Omega_{\textrm{c}}$ values of $0.8\gamma_{41}$, $\gamma_{41}$, and $1.2\gamma_{41}$, respectively. In plots (b) and (c), the mirror frequency is $\omega_{\textrm{m}}=2$ (2$\pi$ MHz). The remaining system parameters remain consistent with those in Fig. \eqref{proberesponse_HAB}.} \label{maxbright1}.
  \end{figure}

In this subsection, we explore the effects of pump and control field strengths on the normalized brightness of the emitted probe field. Figure \ref{maxbright1}(a) illustrates the spectral normalized brightness $\frac{\textrm{B}(\Delta_{\textrm{pr}})}{\textrm{n}_{41}}$ as a function of probe detuning ($\Delta_{\textrm{pr}}$) for pump Rabi frequencies $\Omega_{\textrm{pu}}= 0.8\gamma_{41}$ (solid red curve), $\Omega_{\textrm{pu}}= \gamma_{41}$ (solid black curve), and $\Omega_{\textrm{pu}}= 1.2\gamma_{41}$ (dashed blue curve) in the context of the $\textrm{HE}_{\textrm{pu}}$ engine. The profiles of normalized brightness clearly indicate that with an increase in pump strength, the brightness increases, and the width of the brightness profile also increases. The reason behind this phenomenon lies in the increasing strength of the pump field, which nullifies probe absorption at the line center ($\Delta_{\textrm{pr}}=0$), thereby increasing the amplitude of the EIT window. Simultaneously, it induces power broadening in the absorption profiles—an established concept in EIT \cite{hossain2009study}. Consequently, this results in a brighter emission of light under EIT conditions as the pump strength increases. The emission profiles are not as significant, given their order of magnitude approximately 100 times lesser than the absorption order. Therefore, brightness is predominantly determined by the absorption profiles. As the pump field strength increases, reducing absorption at the line center due to EIT, the emitted light becomes brighter, and the width increases in proportion to the strength of the pump field, reflecting the power broadening effect in the absorption profiles. In Figure \ref{maxbright1}(b), a similar pattern is observed, demonstrating a comparable variation for the $\textrm{HE}_{\textrm{c}}$ engine concerning control field strengths for fixed mirror frequency $\omega_{\textrm{m}}=2$ (2$\pi$ MHz). However, in this scenario, the $\frac{\textrm{B}(\Delta_{\textrm{pr}})}{\textrm{n}_{41}}$ increases gradually with the control field strength $\Omega_{\textrm{c}}$ due to the mirror frequency $\omega_{\textrm{m}}$. Consequently, the presence of mirror vibration impedes the $\textrm{HE}_{\textrm{c}}$ engine from reaching the peak of spectral brightness.

In Fig. \ref{maxbright1}(c), the variation of $\frac{\textrm{B}(\Delta_{\textrm{pr}})}{\textrm{n}_{41}}$ is depicted concerning the probe detuning ($\Delta_{\textrm{pr}}$) for different control Rabi frequencies: $\Omega_{\textrm{c}}= 0.8\gamma_{41}$ (solid red curve), $\Omega_{\textrm{c}}= \gamma_{41}$ (solid black curve), and $\Omega_{\textrm{c}}= 1.2\gamma_{41}$ (dashed blue curve) within the context of the $\textrm{HE}_{\textrm{pu,c}}$ engine. In contrast to the $\textrm{HE}_{\textrm{c}}$ engine, here, the normalized brightness decreases as the strength of the control field increases. The increase in control field strength amplifies the two absorption peaks [as observed in Figure \ref{proberesponse_HAB}(a)] within the probe absorption profiles for the $\textrm{HE}_{\textrm{pu,c}}$ engine. Therefore, it diminishes the brightness of the emitted probe field in Fig. \ref{maxbright1}(c).

\begin{figure}
\centering
\includegraphics[scale=0.55]{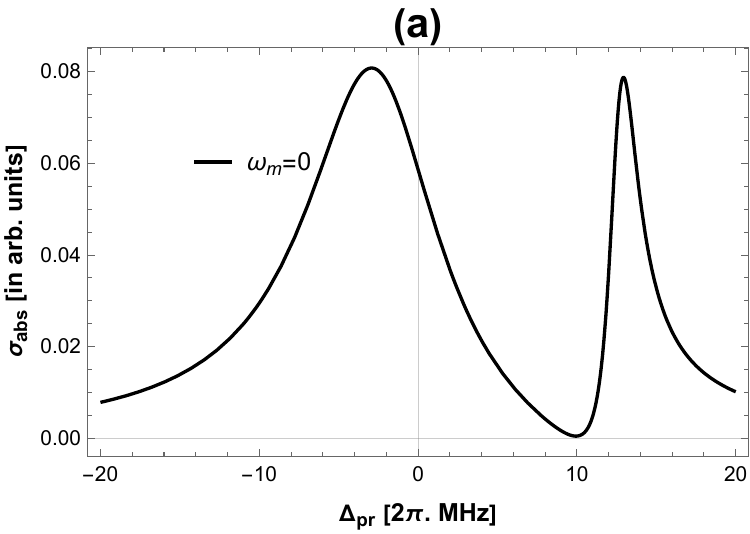}
\includegraphics[scale=0.55]{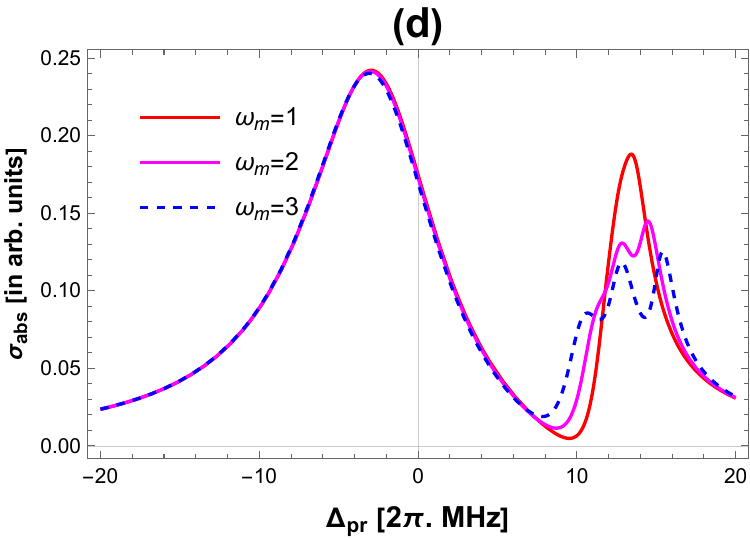}
\includegraphics[scale=0.55]{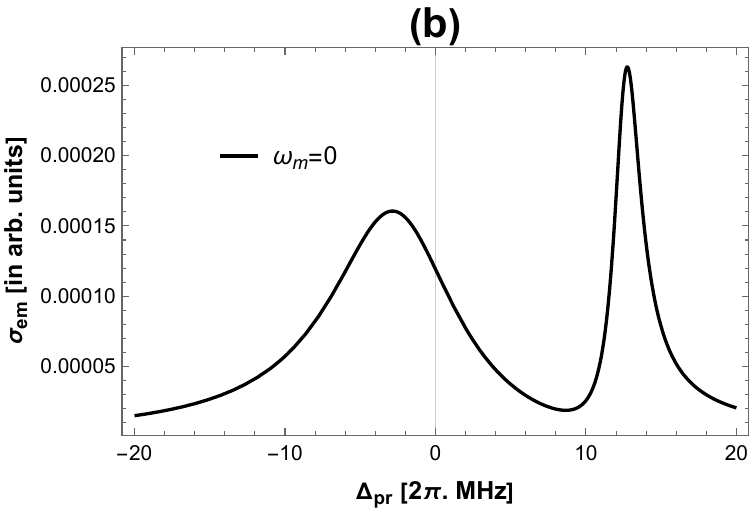}
\includegraphics[scale=0.55]{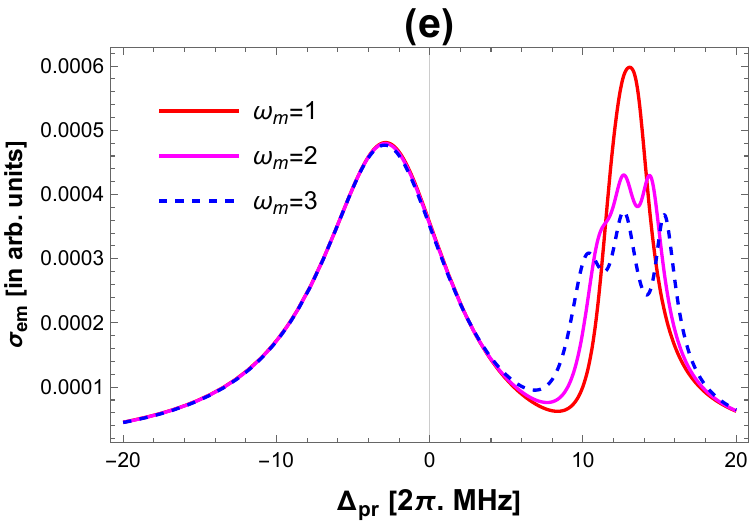}
\includegraphics[scale=0.55]{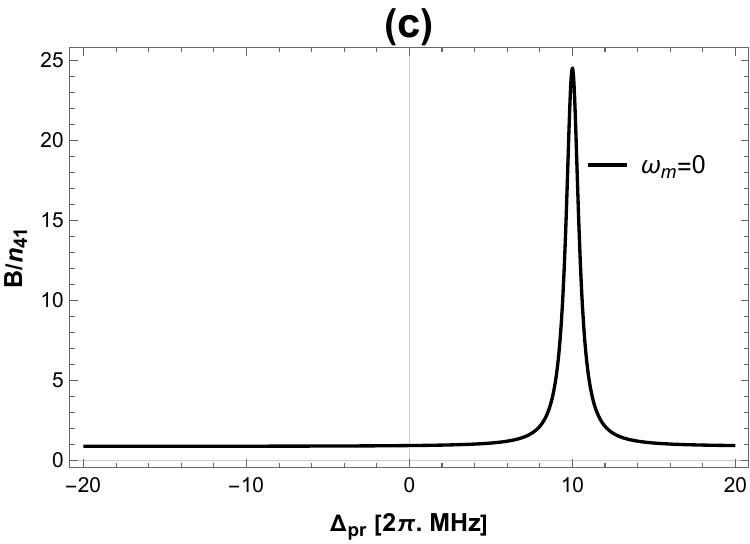}
\includegraphics[scale=0.55]{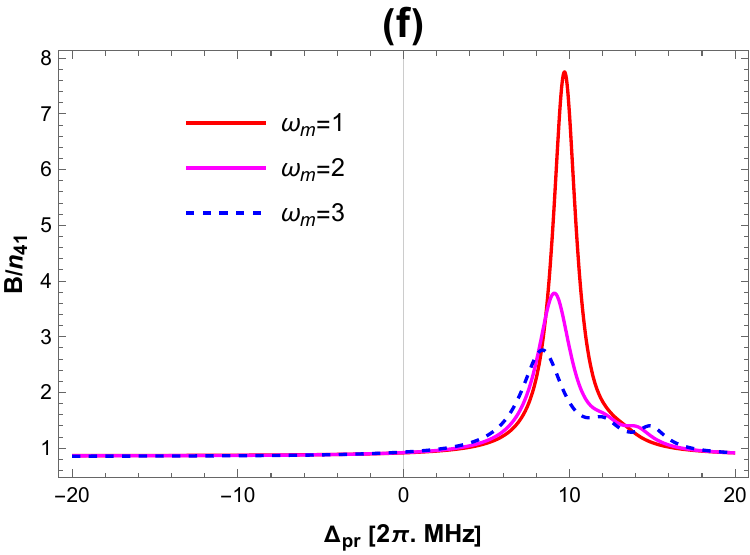}
\caption{The probe absorption coefficient $\sigma_{\textrm{abs}}$ (depicted in plots (a) and (d)), emission coefficient $\sigma_{\textrm{em}}$ (shown in plots (b) and (e)), and normalized brightness $\frac{\textrm{B}(\Delta_{\textrm{pr}})}{\textrm{n}_{41}}$ (illustrated in plots (c) and (f)) are plotted as functions of probe detuning ($\Delta_{\textrm{pr}}$) for \textrm{HE}$_{\textrm{pu}}$ (depicted in the left panel-\textit{i.e.}, (a), (b), and (c)) and \textrm{HE}$_{\textrm{c}}$ (displayed in the right panel-\textit{i.e.}, (c), (e), and (f)) engines. The solid black curve (in the left panel), solid red curve (in the right panel), solid magenta curve (in the right panel), and dashed blue curve (in the right panel) represent mirror frequency $\omega_{\textrm{m}}=0$, $\omega_{\textrm{m}}=1$ (2$\pi$. MHz), $\omega_{\textrm{m}}=2$ (2$\pi$. MHz), and $\omega_{\textrm{m}}=3$ (2$\pi$. MHz) respectively. In the left panel, $\Delta_{\textrm{pu}}=10$ (2$\pi$. MHz), and in the right panel, $\Delta_{\textrm{c}}=10$ (2$\pi$. MHz). The remaining system parameters match those in Fig. \eqref{proberesponse}.}\label{control1}
\end{figure}

Figs. \ref{control1}(a), \ref{control1}(b), and \ref{control1}(c) respectively illustrate the absorption coefficient ($\sigma_{\textrm{abs}}$), the emission coefficient ($\sigma_{\textrm{em}}$), and the normalized spectral brightness $\frac{\textrm{B}(\Delta_{\textrm{pr}})}{\textrm{n}_{41}}$ as functions of probe detuning ($\Delta_{\textrm{pr}}$) for $\Delta_{\textrm{pu}}=10$ (2$\pi$ MHz) within the $\textrm{HE}_{\textrm{pu}}$ engine. Here, the pump detuning not only shifts the position of brightness at $\Delta_{\textrm{pr}}-\Delta_{\textrm{pu}}=0$ due to the Raman resonance condition but also reduces the normalized brightness of the emitted probe light, as seen in Figs. \ref{control1}(c) and \ref{proberesponse}(c) (when $\Delta_{\textrm{pu}}=0$). The reason is that as the pump detuning increases and reaches the width of the Lorentzian profile of the population of level $\ket{4}$ curve against probe detuning, \textit{i.e.}, $\rho_{44}$ versus $\Delta_{\textrm{pr}}$ Lorentzian profile, the atoms in level $\ket{2}$ do not participate in the probe emission process \cite{Liu_2020}. This leads to a reduction in brightness. A similar pattern is observed in the right panel of Figure \ref{control1} for the $\textrm{HE}_{\textrm{c}}$ engine at a control detuning of $\Delta_{\textrm{c}}=10$ (2$\pi$ MHz). Here, the normalized brightness shifts according to the Raman resonance condition, \textit{i.e.}, $\Delta_{\textrm{pr}}-\Delta_{\textrm{c}}+\frac{\omega_{m}}{2}=0$, as both positive and negative mirror frequencies, \textit{i.e.}, $\Tilde{\rho}_{jk, \pm}$, are considered in the calculation.

\begin{figure}
\centering
\includegraphics[scale=0.55]{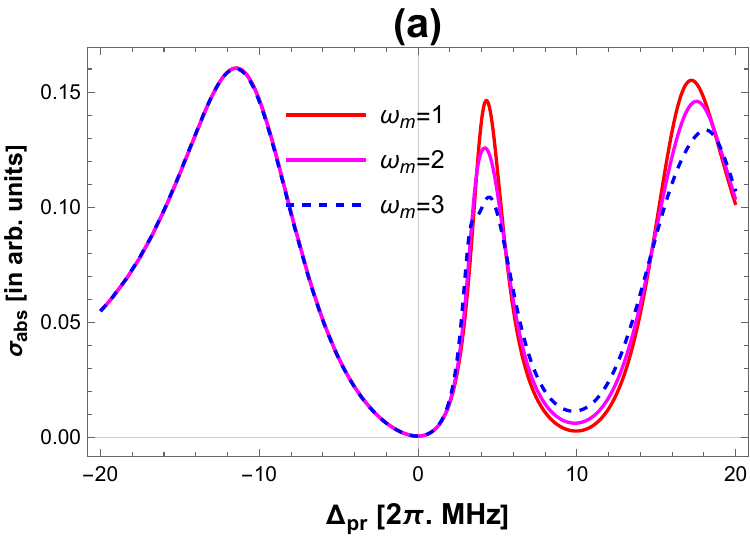}
\includegraphics[scale=0.55]{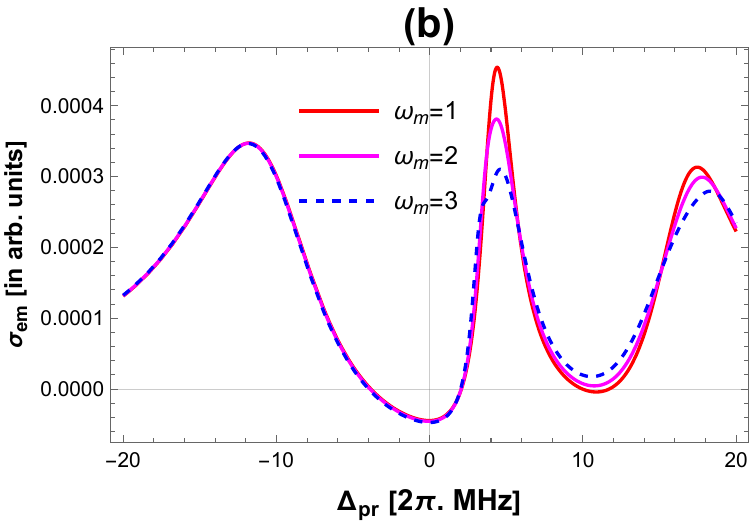}
\includegraphics[scale=0.55]{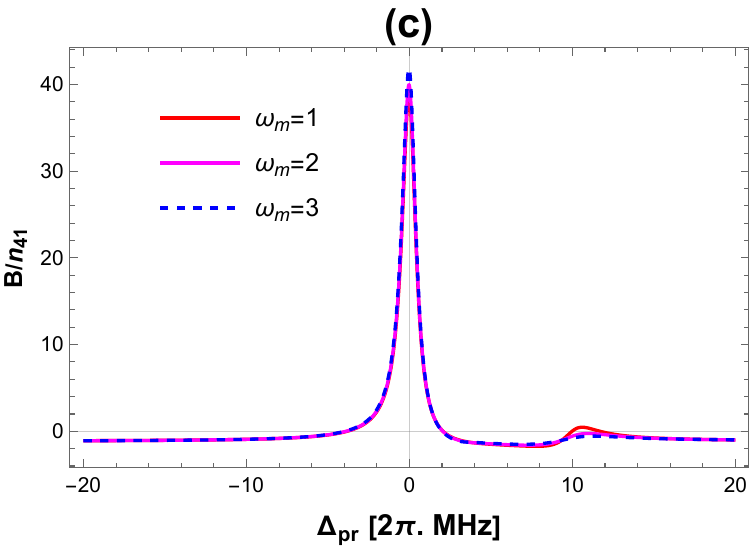}
\caption{The probe absorption coefficient $\sigma_{\textrm{abs}}$ [in the plot (a)], emission coefficient $\sigma_{\textrm{em}}$  [in the plot (b)] and normalized brightness $\frac{\textrm{B}(\Delta_{\textrm{pr}})}{\textrm{n}_{41}}$  [in the plot (c)] are depicted as a function of probe detuning ($\Delta_{\textrm{pr}}$) for composite \textrm{HE}$_{\textrm{pu,c}}$ engine. The control detiuning is $\Delta_{\textrm{c}}=10$ (2$\pi$. MHz). The solid red color, solid magenta color, and dashed blue color curves are for    $\omega_{\textrm{m}}=1$ (2$\pi$. MHz),  $\omega_{\textrm{m}}=2$ (2$\pi$. MHz), and  $\omega_{\textrm{m}}=3$ (2$\pi$. MHz), respectively. The remaining system parameters match those in Fig.\eqref{proberesponse_HAB}} \label{control2}
\end{figure}

Fig. \ref{control2} illustrates the variations of the probe absorption coefficient $\sigma_{\textrm{abs}}$ (Fig. \ref{control2}(a)), emission coefficient $\sigma_{\textrm{em}}$ (Fig. \ref{control2}(b)), and the normalized spectral brightness $\frac{\textrm{B}(\Delta_{\textrm{pr}})}{\textrm{n}_{41}}$ (Fig. \ref{control2}(c)) concerning the probe detuning $\Delta_{\textrm{pr}}$, while maintaining a control detuning of $\Delta_{\textrm{c}}=10$ (2$\pi$ MHz) for $\textrm{HE}_{\textrm{pu,c}}$ engine. Double EIT windows are observed in both absorption and emission profiles at $\Delta_{\textrm{pr}}=0$ and $\Delta_{\textrm{pr}}-\Delta_{\textrm{c}}=0$. The mirror frequency affects only at $\Delta_{\textrm{pr}}-\Delta_{\textrm{c}}=0$, where it increases the EIT window in both absorption and emission profiles. The occurrence of these double EIT windows is attributed to the cross-phase interaction between the emitted probe and the applied control field, which is comprehensively discussed in reference \cite{PhysRevA.89.021802} using a four-level tripod-type system. In Fig. \ref{control2}(c), the normalized brightness shows almost no peak at $\Delta_{\textrm{pr}}-\Delta_{\textrm{c}}=0$. Additionally, the control detuning $\Delta_{\textrm{c}}$ has minimal effect in increasing the brightness at $\Delta_{\textrm{pr}}=0$, contrary to its effects on the $\textrm{HE}_{\textrm{c}}$ engine.

\subsection{ Amplitude and phase of modulated emitted probe coherence term}

\begin{figure}
\centering
\includegraphics[scale=0.55]{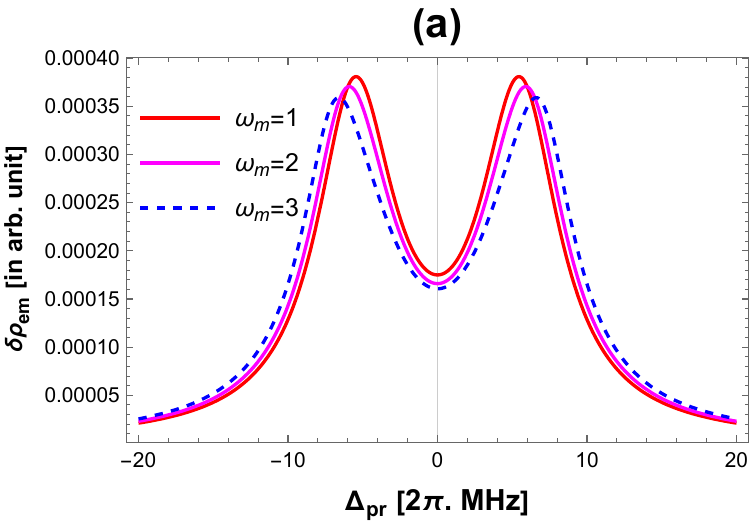}
\includegraphics[scale=0.55]{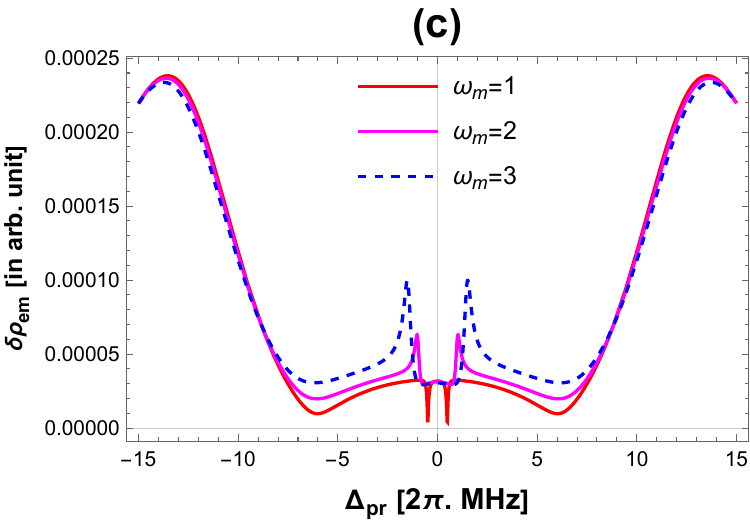}
\includegraphics[scale=0.55]{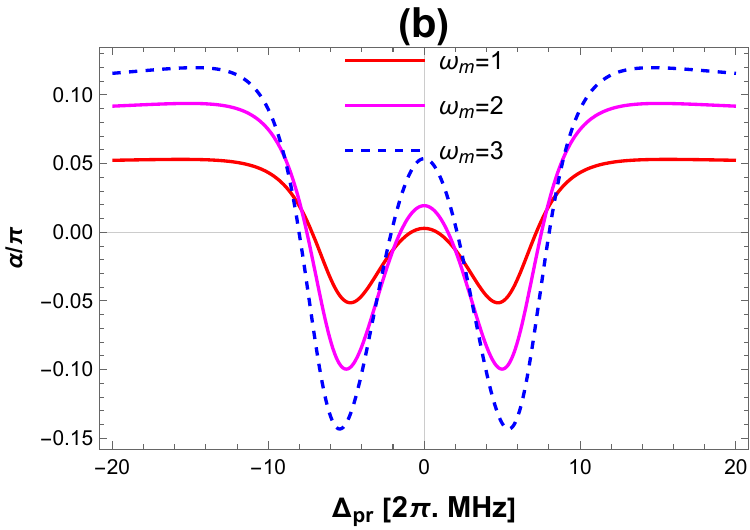}
\includegraphics[scale=0.55]{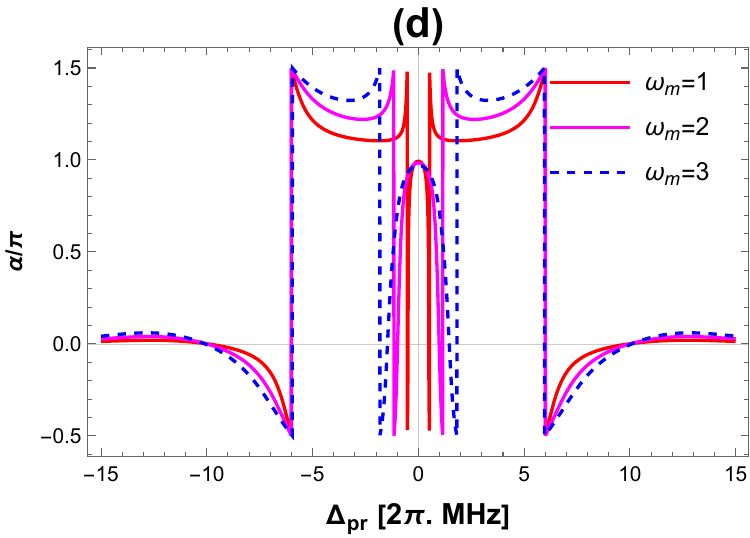}
\caption{The amplitude of the modulated probe emission term $\delta\Tilde{\rho}_{14}^{em}$ is varied against probe detuning $\Delta_{\textrm{pr}}$ for (a) $\textrm{HE}_{\textrm{c}}$ and (c) $\textrm{HE}_{\textrm{pu, c}}$ engines. The phase $\alpha/\pi$ between probe emission modulation and mirror motion is depicted as a function of $\Delta_{\textrm{pr}}$ for (b) $\textrm{HE}_{\textrm{c}}$ and (d) $\textrm{HE}_{\textrm{pu, c}}$ engines. The solid red color, solid magenta color, and dashed blue color curves are for    $\omega_{\textrm{m}}=1$ (2$\pi$. MHz),  $\omega_{\textrm{m}}=2$ (2$\pi$. MHz), and  $\omega_{\textrm{m}}=3$ (2$\pi$. MHz), respectively. The other system parameters are identical to those in Fig. \eqref{proberesponse_HAB}\label{modulation1}}
\end{figure}

In Figure \ref{modulation1}(a), the amplitude of the oscillatory emitted probe field coherence, $\delta\Tilde{\rho}_{14,0}^{em}$, is depicted as a function of the probe detuning $\Delta_{\textrm{pr}}$ for the $\textrm{HE}_{\textrm{c}}$ heat engine. Notably, $\delta\Tilde{\rho}_{14,0}^{em}$ exhibits a prominent reduction window near $\Delta_{\textrm{pr}}=0$, indicating a substantial decrease in the amplitude of oscillations in the imaginary part of the probe transition coherence, $\textrm{Im}[\Tilde{\rho}_{14, 0}^{em}]$. This reduction occurs when the EIT condition is fulfilled, and with increasing mirror frequency, these amplitudes are further reduced at $\Delta_{\textrm{pr}}=0$. The amplitude of the oscillatory emitted probe field coherence should typically increase with the mirror frequency. In other words, the more the nano mirror vibrates, the greater the amplitude of the oscillatory emitted probe field induced by the mirror's vibration. However, in the presence of EIT, the system behaves oppositely \textit{i.e.} the higher the mirror frequency of the nanomechanical vibration, the more the amplitude is reduced. In general, the constant mirror vibration induces oscillation in the density matrix element and consequently, a phase difference arises between the oscillation of the amplitude of the density matrix element and the mirror motion. It is expected that as the mirror vibrates more, a greater phase difference will manifest. Figure \ref{modulation1}(b) illustrates the phase difference $\alpha$ between the modulation of probe emission and the motion of the mirror in the context of the $\textrm{HE}_{\textrm{c}}$ engine. At around $\Delta_{\textrm{pr}}=0$, the phase difference $\alpha$ increases with mirror frequency, indicating that as the mirror frequency rises in the $\textrm{HE}_{\textrm{c}}$ engine, so does the phase difference between the modulation of probe emission and the motion of the mirror.

Figure \ref{modulation1}(c) illustrates the plot of the modulated emitted probe field amplitude, $\delta\Tilde{\rho}_{14,0}^{em}$, as a function of the probe detuning $\Delta_{\textrm{pr}}$ for the $\textrm{HE}_{\textrm{pu,c}}$ heat engine. Notably, $\delta\Tilde{\rho}_{14,0}^{em}$ shows no changes concerning mirror frequency around $\Delta_{\textrm{pr}}=0$, contradicting the behavior expected in the context of the $\textrm{HE}_{\textrm{c}}$ engine. Although, $\delta\Tilde{\rho}_{14,0}^{em}$ decreases at $\Delta_{\textrm{pr}}=0$ but mirror frequency does not affect that much as it does in the $\textrm{HE}_{\textrm{c}}$ engine. Figure \ref{modulation1}(d) illustrates the phase difference $\alpha$ between the modulation of probe emission and the motion of the mirror in the context of the $\textrm{HE}_{\textrm{pu,c}}$ heat engine. A distinct peak emerges near $\Delta_{\textrm{pr}}=0$ for the phase difference $\alpha$, maintaining a consistent amplitude regardless of the mirror frequency while exhibiting an increased width with higher mirror vibration. Consequently, the phase difference remains unaffected by the mirror frequency. However, the range of emitted probe frequencies exhibiting a phase shift between the modulation of probe emission and the motion of the mirror increases for the $\textrm{HE}_{\textrm{pu,c}}$ heat engine.

\subsection{Entropy and Second Law of Thermodynamics}

In the proposed heat engine model, in the absence of laser fields, the spectral brightness $\textrm{B}(\Delta_{\textrm{pr}})$ is related to the photon no. $\textrm{n}_{41}$ of the $\textrm{T}_{41}$ reservoir through detailed balance, and it can be expressed as $\textrm{B}(\Delta_{\textrm{pr}})=\textrm{n}_{41}$ and temperature of the $\textrm{T}_{41}$ reservoir attains its maximum value \textit{i.e.} 
\begin{eqnarray}\label{temp_max}
   \textrm{T}_{\textrm{max}}=\frac{\hbar\omega_{41}}{\textrm{k}_{\textrm{B}}[ln(\frac{1}{\textrm{B}(\Delta_{\textrm{pr}})}+1)]}. 
\end{eqnarray}
The maximum temperature ($\textrm{T}_{\textrm{max}}$), considering pump and control fields, is limited by the requirement of increasing entropy \cite{PhysRevA.94.053859}, \textit{i.e.}

\begin{eqnarray}
    \begin{aligned}
        \Delta\textrm{S}&=-\frac{\hbar\omega_{41}}{\textrm{T}_{41}}+\frac{\hbar\omega_{42}}{\textrm{T}_{42}}+\frac{\hbar\omega_{43}}{\textrm{T}_{43}}+\frac{\hbar\omega_{41}}{\textrm{T}_{\textrm{B}}}\geq 0~~\textrm{for}~~\textrm{HE}_{\textrm{pu,c}}~~\textrm{heat engine}\\
         \Delta\textrm{S}&=-\frac{\hbar\omega_{41}}{\textrm{T}_{41}}+\frac{\hbar\omega_{42}}{\textrm{T}_{42}}+\frac{\hbar\omega_{41}}{\textrm{T}_{\textrm{B}}}\geq 0~~\textrm{for}~~\textrm{HE}_{\textrm{pu}}~~\textrm{heat engine}\\
          \Delta\textrm{S}&=-\frac{\hbar\omega_{41}}{\textrm{T}_{41}}+\frac{\hbar\omega_{43}}{\textrm{T}_{43}}+\frac{\hbar\omega_{41}}{\textrm{T}_{\textrm{B}}}\geq 0~~\textrm{for}~~\textrm{HE}_{\textrm{c}}~~\textrm{heat engine}\\
    \end{aligned}
\end{eqnarray}

Here, the entropy of emitted radiation with temperature $\textrm{T}_{\textrm{B}}$ can be written as $\textrm{S}=\frac{\hbar\omega_{41}}{\textrm{T}_{\textrm{B}}}$. For $\textrm{T}_{41}=\textrm{T}_{42}=\textrm{T}_{43}=\textrm{T}_{0}=5000$K, the limits of temperature for emitted probe fields from the entropy balance conditions are such that $\frac{\textrm{T}_{\textrm{B}}}{\textrm{T}_{0}} \geq 2$ for $\textrm{HE}_{\textrm{pu,c}}$, $\frac{\textrm{T}_{\textrm{B}}}{\textrm{T}_{0}} \leq 4$ for $\textrm{HE}_{\textrm{pu}}$, and $\frac{\textrm{T}_{\textrm{B}}}{\textrm{T}_{0}} \leq 4$ for $\textrm{HE}_{\textrm{c}}$.

Table \ref{tab_pu} illustrates the impact of the pump laser field on the scaled maximum temperature $\frac{\textrm{T}_{\textrm{max}}}{\textrm{T}_{0}}$ for the $\textrm{HE}_{\textrm{pu}}$ engine. As the pump field is increased from 0 to $2.5\gamma_{41}$, $\frac{\textrm{T}{\textrm{max}}}{\textrm{T}_{0}}$ rises from 1.438 to 3.884. Table \ref{tab_c} illustrates how the control laser field impacts the scaled maximum temperature, denoted as $\frac{\textrm{T}_{\textrm{max}}}{\textrm{T}_{0}}$, in the context of the $\textrm{HE}_{\text{c}}$ engine. The effects are analyzed concerning different mirror frequencies: $\omega_{\textrm{m}}=1$ ($2\pi$ MHz), $\omega_{\textrm{m}}=2$ ($2\pi$ MHz), and $\omega_{\textrm{m}}=3$ ($2\pi$ MHz). For $\omega_{\textrm{m}}=1$ ($2\pi$ MHz), the ratio $\frac{\textrm{T}_{\textrm{max}}}{\textrm{T}_{0}}$ shows an increase from 1.438 to 2.535 as the control field varies from 0 to $2\gamma_{41}$. Similarly, for $\omega_{\textrm{m}}=2$ ($2\pi$ MHz), the ratio increases from 1.438 to  2.186, and for $\omega_{\textrm{m}}=3$ ($2\pi$ MHz), it rises from 1.438 to 2.050 under the same conditions. This demonstrates a gradual increase in $\frac{\textrm{T}_{\textrm{max}}}{\textrm{T}_{0}}$ with rising $\omega_{\textrm{m}}$, albeit at a slower pace.

The maximum temperatures observed in both engines consistently comply with the upper limit set by the entropy balance. Notably, the tables (Table \ref{tab_pu} and Table \ref{tab_c}) reveal a consistent trend: the scaled maximum temperature decreases with mirror vibration. This phenomenon indicates that mirror vibration causes the heat engine (three-level $\Lambda$-type engine) to deviate from the ideal state of a heat engine. 

Table \ref{tab_pu,c} offers a detailed comparison of how the control field influences the scaled maximum temperature $\frac{\textrm{T}_{\textrm{max}}}{\textrm{T}_{0}}$ in the $\textrm{HE}_{\textrm{pu,c}}$ engine under varying mirror vibrations. For $\omega_{\textrm{m}}=1$ ($2\pi$ MHz), the $\frac{\textrm{T}_{\textrm{max}}}{\textrm{T}_{0}}$ increases from 2.589 to 2.825 with the control field ranging from 0 to $2\gamma_{41}$. However, at $\omega_{\textrm{m}}=2$ ($2\pi$ MHz) and $\omega_{\textrm{m}}=3$ ($2\pi$ MHz), the $\frac{\textrm{T}_{\textrm{max}}}{\textrm{T}_{0}}$ rises from 2.589 to 3.018 and 2.589 to 3.188, respectively, as the control field undergoes the same increase. The presence of mirror vibration in the $\textrm{HE}_{\textrm{pu,c}}$ engine results in a greater increase in the scaled maximum temperature $\frac{\textrm{T}_{\textrm{max}}}{\textrm{T}_{0}}$ compared to the $\textrm{HE}_{\textrm{c}}$ engine.

The entropy flow rate per unit power, also known as the reciprocal of the radiation's flux temperature $\textrm{T}_{\textrm{B}}$ \cite{PhysRevA.103.062205,PhysRevB.75.214304, doi:10.1119/1.1842732}, is obtained by considering the brightness $\textrm{B}(\Delta_{\textrm{pr}})$ as the photon occupation number \textit{i.e.}
\begin{eqnarray}\label{entropy}
    \frac{\textrm{S}}{\textrm{k}_{\textrm{B}}}= \frac{\hbar\omega_{41}}{\textrm{T}_{\textrm{B}}}=\frac{\int[(\textrm{B}+1)ln(\textrm{B}+1)-\textrm{B}ln\textrm{B}]d\omega_{\textrm{pr}}}{\int \textrm{B} d\omega_{\textrm{pr}}}.
\end{eqnarray}
To proceed with the discussion on entropy ($\frac{\textrm{S}}{\textrm{k}_{\textrm{B}}}$) variation concerning applied laser field strengths, it is essential to define the upper and lower bounds for our system's entropy, considering the requirement of increasing entropy (\cite{PhysRevA.103.062205}). The upper limit for all suggested heat engines is determined by the entropy balance condition, \textit{i.e.}, 
\begin{eqnarray}
    \textrm{S}-\frac{\hbar\omega_{41}}{\textrm{T}_{41}}\leq 0
\end{eqnarray}
In the context of the upper limit, the situation encompasses the absence of applied laser fields (pump and control) and the $\textrm{T}_{42}$, $\textrm{T}_{43}$ reservoirs. The lower limit is determined by the entropy balance condition, \textit{i.e.}  
\begin{eqnarray}
\begin{aligned}
    \textrm{S}&-\frac{\hbar\omega_{41}}{\textrm{T}_{41}}+\frac{\hbar\omega_{42}}{\textrm{T}_{42}}+\frac{\hbar\omega_{43}}{\textrm{T}_{43}}\geq 0 ~~\textrm{for}~~\textrm{HE}_{\textrm{pu,c}}~~\textrm{heat engine}\\
     \textrm{S}&-\frac{\hbar\omega_{41}}{\textrm{T}_{41}}+\frac{\hbar\omega_{42}}{\textrm{T}_{42}}\geq 0~~\textrm{for}~~\textrm{HE}_{\textrm{pu}}~~\textrm{heat engine}\\
      \textrm{S}&-\frac{\hbar\omega_{41}}{\textrm{T}_{41}}+\frac{\hbar\omega_{43}}{\textrm{T}_{43}}\geq 0~~\textrm{for}~~\textrm{HE}_{\textrm{c}}~~\textrm{heat engine}
\end{aligned}
    \end{eqnarray}
The upper and lower limits of entropy for $\textrm{HE}_{\textrm{pu}}$ and $\textrm{HE}_{\textrm{c}}$ engines are 6.110 and 1.527, respectively. Likewise, the $\textrm{HE}_{\textrm{pu,c}}$ engine has upper and lower limits of 6.110 and -3.05, respectively. In the three heat engines (Tables \ref{tab_pu}, \ref{tab_c}, and \ref{tab_pu,c}), as the strength of the pump field increases (for $\textrm{HE}_{\textrm{pu}}$ engine) or the control field increases (for $\textrm{HE}_{\textrm{c}}$ engine), entropy decreases. However, in the $\textrm{HE}_{\textrm{pu, c}}$ engine, entropy slightly increases with the control field strength in the presence of mirror frequency. When mirror vibration is present, the entropy decrement in the $\textrm{HE}_{\textrm{c}}$ engine is less pronounced, whereas the slight increment in entropy in the $\textrm{HE}_{\textrm{pu, c}}$ engine slightly increases in the presence of mirror frequency.

To comprehend the entropy profile variation caused by mirror vibration, the emission rate $\textrm{R}$ is estimated. The emission rate is inversely proportional to the entropy flow rate per unit power of the emitted probe field. This rate represents the total number of output photons generated per second and is defined as,
\begin{eqnarray}\label{emission}
    \textrm{R}=\frac{1}{2\pi}\int \textrm{B}(\Delta_{\textrm{pr}})d\omega_{\textrm{pr}}
\end{eqnarray}
 As the strength of the pump field increases in the $\textrm{HE}_{\textrm{pu}}$ engine (Table \ref{tab_pu}), the emission rate also increases. However, in the case of the $\textrm{HE}_{\textrm{c}}$ engine, the emission rate increases slowly with the control field strength. Consequently, mirror vibration plays a crucial role in controlling the emission rate of the $\textrm{HE}_{\textrm{c}}$ engine (Table \ref{tab_c}). However, in $\textrm{HE}_{\textrm{pu,c}}$, the emission rate slightly decreases with the increasing strength of the control field (Table \ref{tab_pu,c}).

\section{Conclusions}\label{con}
The paper presents a theoretical model of three different non-traditional heat engines utilizing a multi-level tripod-type atomic system coupled with mirror vibration and black body reservoirs. The model yields intriguing aspects that can be highlighted as follows:
\begin{itemize}
\item The model integrates mirror vibration while disregarding cavity confinement, establishing a connection between a multi-level atomic system and the mechanical vibration of the mirror, functioning as heat engines. These heat engines include three-level $\Lambda$-type engines, such as $\textrm{HE}_{\textrm{pu}}$ and $\textrm{HE}_{\textrm{c}}$, as well as four-level tripod-type engines like $\textrm{HE}_{\textrm{pu,c}}$.
 \item The heat engine $\textrm{HE}_{\textrm{pu}}$ exhibits the highest normalized brightness, while the composite heat engine $\textrm{HE}_{\textrm{pu,c}}$ displays the second-highest value. This brightness experiences a slight increase with rising mirror frequency. On the contrary, in $\textrm{HE}_{\textrm{c}}$, the normalized brightness reaches its minimum and decreases as the mirror frequency rises.
 \item The pump field increases the spectral normalized brightness of the $\textrm{HE}_{\textrm{pu}}$ engine, and a comparable effect is observed in the $\textrm{HE}_{\textrm{c}}$ engine with the control field. However, in the $\textrm{HE}_{\textrm{pu,c}}$ engine, the normalized brightness of the emitted probe field decreases as the control field's value increases.

 \item Pump detuning (or control detuning) not only shifts the position of the normalized brightness of emitted probe light in the $\textrm{HE}_{\textrm{pu,c}}$ (or $\textrm{HE}_{\textrm{c}}$) engine but also reduces the normalized brightness. However, it hardly affects the $\textrm{HE}_{\textrm{pu,c}}$ engine.
 \item The amplitude of the oscillatory emitted probe field coherence decreases with mirror frequency at the line center of the emitted probe field due to the EIT condition in both $\textrm{HE}_{\textrm{c}}$ and $\textrm{HE}_{\textrm{pu,c}}$ heat engines. Conversely, the phase difference between the mirror motion and amplitude increases with mirror frequency for the $\textrm{HE}_{\textrm{c}}$ engine, whereas it remains nearly constant at the line center of the $\textrm{HE}_{\textrm{pu,c}}$ heat engine.
 \item Mirror vibration causes both the $\textrm{HE}_{\textrm{c}}$ and $\textrm{HE}_{\textrm{pu,c}}$ engines to deviate from their ideal heat engine behavior. As respective pump and control field strength increases, the entropy flow rate decreases in both $\textrm{HE}_{\textrm{pu}}$ and $\textrm{HE}_{\textrm{c}}$ engines. In the $\textrm{HE}_{\textrm{pu, c}}$ engine, there is a slight increase in entropy as the strength of the control field grows in the presence of mirror frequency.
 \item Stronger pump fields in $\textrm{HE}_{\textrm{pu}}$  correspond to increased emission rates. Contrastingly, while the $\textrm{HE}_{\textrm{c}}$ engine shows a slow rise in emission rate with control field strength, mirror vibration notably influences and controls the emission rate in this engine. In $\textrm{HE}_{\textrm{pu,c}}$, the emission rate exhibits a slight decrease as the strength of the control field increases.
\end{itemize}

Hence, this simple theoretical model assists in comprehending the impact of mirror vibration on the photons emitted by a heat engine within a semi-classical framework, without resorting to a complete quantum mechanical treatment.

\section*{Appendix I}\label{appi}
The total electric field due to both the applied pump and control fields, as well as the emitted probe field, can be written as
\begin{eqnarray}
    \mathbf{E}=\mathcal{\mathbf{E}}_{pr}\textrm{cos}(\omega_{pr}t-\mathbf{k}_{pr}.\mathbf{r})+\mathcal{\mathbf{E}}_{pu}\textrm{cos}(\omega_{pu}t-\mathbf{k}_{pu}.\mathbf{r})+\mathcal{\mathbf{E}}_{c}\textrm{cos}\lbrace(\omega_{c}\pm \omega_{m})t-\mathbf{k}_{c}.\mathbf{r}\rbrace
\end{eqnarray}
The electric field amplitudes, frequencies, and wave vectors for the probe (pump and control) are denoted as $\mathcal{\mathbf{E}}_{pr}$ ($\mathcal{\mathbf{E}}_{pu}$ and $\mathcal{\mathbf{E}}_{c}$), $\omega_{pr}$ ($\omega_{pu}$ and $\omega_{c}$), and $\mathbf{k}_{pr}$ ($\mathbf{k}_{pu}$ and $\mathbf{k}_{c}$), respectively. The parameters $\omega_{m}$, $\mathbf{r}$, and $t$ correspond to the mirror's frequency, atomic position, and time, respectively. Neglecting $\mathbf{k}.\mathbf{r}$ due to the larger wavelength of the laser fields compared to the atom's diameter (dipole approximation), the electric field can be succinctly represented as
\begin{eqnarray}\label{Electric}
\mathbf{E}=\mathcal{\mathbf{E}}_{pr}\textrm{cos}(\omega_{pr}t)+\mathcal{\mathbf{E}}_{pu}\textrm{cos}(\omega_{pu}t)+\mathcal{\mathbf{E}}_{c}\textrm{cos}\lbrace(\omega_{c}\pm \omega_{m})t\rbrace
\end{eqnarray}
The total Hamiltonian of the system is now expressed as, 
\begin{eqnarray}
\mathcal{H}=\mathcal{H}_{0}+\mathcal{H}_{I}
\end{eqnarray}
The bare atom Hamiltonian ($\mathcal{H}_{0}$) is given by
\begin{eqnarray}
\mathcal{H}_{0}=\begin{pmatrix}
\hbar\omega_{1}&0&0&0\\
0&\hbar\omega_{2}&0&0\\
0&0&\hbar\omega_{3}&0\\
0&0&0&\hbar\omega_{4}
\end{pmatrix}=\hbar(\omega_{1}\sigma_{11}+\omega_{2}\sigma_{22}+\omega_{3}\sigma_{33}+\omega_{4}\sigma_{44})
\end{eqnarray}
The characteristic frequency of level $\ket{i}$ is denoted as $\omega_{i}$ for $i=1, 2, 3, 4$, and $\sigma_{jj}=\ket{j}\bra{j}$ represents the projection operator. The interaction Hamiltonian ($\mathcal{H}_{I}$) is defined as
\begin{eqnarray}
\mathcal{H}_{I}=-\mathcal{\mathbf{E}}\begin{pmatrix}
0&0&0&\mathbf{d}_{14}\\
0&0&0&\mathbf{d}_{24}\\
0&0&0&\mathbf{d}_{34}\\
\mathbf{d}_{14}&\mathbf{d}_{24}&\mathbf{d}_{34}&0
\end{pmatrix}
\end{eqnarray}
The dipole matrix element $\mathbf{d}_{jk}=\bra{j}\mathbf{d}\ket{k}$ is real, implying $\mathbf{d}_{jk}=\mathbf{d}_{kj}$. Transitions $\ket{1}\rightarrow\ket{2}$, $\ket{2}\rightarrow\ket{3}$, and $\ket{1}\rightarrow\ket{3}$ are dipole forbidden. The zero values of the diagonal elements in $\mathcal{H}_{I}$ stem from the symmetric properties of the wave function. To analyze $\mathbf{E}$'s impact, transform $\mathcal{H}_{I}$ into the unperturbed system's interaction picture using the unitary matrix $\mathcal{U}(t)$ 
\begin{eqnarray}
\mathcal{U}(t)=e^{i\mathcal{H}_{0}t/\hbar}=\begin{pmatrix}
e^{i\omega_{1}t}&0&0&0\\
0&e^{i\omega_{2}t}&0&0\\
0&0&e^{i\omega_{3}t}&0\\
0&0&0&e^{i\omega_{4}t}
\end{pmatrix}
\end{eqnarray}
Employing this transformation on $\mathcal{H}_{I}$ yields
\begin{eqnarray}
\mathcal{U}(t)\mathcal{H}_{I}\mathcal{U}(t)^{\dagger}=-\mathcal{\mathbf{E}}\begin{pmatrix}
0&0&0&\mathbf{d}_{14}e^{-i\omega_{14}t}\\
0&0&0&\mathbf{d}_{24}e^{-i\omega_{24}t}\\
0&0&0&\mathbf{d}_{34}e^{-i\omega_{34}t}\\
\mathbf{d}_{14}e^{i\omega_{14}t}&\mathbf{d}_{24}e^{i\omega_{24}t}&\mathbf{d}_{34}e^{i\omega_{34}t}&0
\end{pmatrix}
\end{eqnarray}
Expressing $\mathbf{E}$ in equation \eqref{Electric} involves a summation of exponentials
\begin{eqnarray}
    \mathbf{E}=\frac{\mathcal{\mathbf{E}}_{pr}}{2}(e^{i\omega_{pr}t}+e^{-i\omega_{pr}t})+\frac{\mathcal{\mathbf{E}}_{pu}}{2}(e^{i\omega_{pu}t}+e^{-i\omega_{pu}t})+\frac{\mathcal{\mathbf{E}}_{c}}{2}(e^{i(\omega_{c}\pm \omega_{m})t}+e^{-i(\omega_{c}\pm \omega_{m})t})
\end{eqnarray}
Substituting this $\mathbf{E}$ into the transformed Hamiltonian and applying the Rotating Wave Approximation allows us to discard rapidly oscillating terms that quickly average out, retaining the slowly oscillating terms. The Hamiltonian ($\mathcal{H}_{I}$) becomes
\begin{eqnarray}
\begin{pmatrix}
0&0&0&-\frac{1}{2}\mathbf{d}_{14}\mathcal{\mathbf{E}}_{pr}e^{i(\omega_{pr}-\omega_{14})t}\\
0&0&0&-\frac{1}{2}\mathbf{d}_{24}\mathcal{\mathbf{E}}_{pu}e^{i(\omega_{pu}-\omega_{24})t}\\
0&0&0&-\frac{1}{2}\mathbf{d}_{34}\mathcal{\mathbf{E}}_{c}e^{i(\omega_{c}-\omega_{34})t}\\
-\frac{1}{2}\mathbf{d}_{14}\mathcal{\mathbf{E}}_{pr}e^{-i(\omega_{pr}-\omega_{14})t}&-\frac{1}{2}\mathbf{d}_{24}\mathcal{\mathbf{E}}_{pu}e^{-i(\omega_{pu}-\omega_{24})t}&-\frac{1}{2}\mathbf{d}_{34}\mathcal{\mathbf{E}}_{c}e^{-i(\omega_{c}-\omega_{34})t}&0
\end{pmatrix}
\end{eqnarray}
Upon reverting to the Schrödinger picture, the interaction Hamiltonian takes the form
\begin{eqnarray}
\begin{aligned}
    \mathcal{H}_{I}=&\mathcal{U}(t)^{\dagger}(\mathcal{U}(t)\mathcal{H}_{I}\mathcal{U}(t)^{\dagger})\mathcal{U}(t)\\=&-\frac{1}{2}\begin{pmatrix}
0&0&0&\mathbf{d}_{14}\mathcal{\mathbf{E}}_{pr}e^{i\omega_{pr}t}\\
0&0&0&\mathbf{d}_{24}\mathcal{\mathbf{E}}_{pu}e^{i\omega_{pu}t}\\
0&0&0&\mathbf{d}_{34}\mathcal{\mathbf{E}}_{c}e^{i(\omega_{c}\pm \omega_{m})t}\\
\mathbf{d}_{14}\mathcal{\mathbf{E}}_{pr}e^{-i\omega_{pr}t}&\mathbf{d}_{24}\mathcal{\mathbf{E}}_{pu}e^{-i\omega_{pu}t}&\mathbf{d}_{34}\mathcal{\mathbf{E}}_{c}e^{-i(\omega_{c}\pm \omega_{m})t}&0
\end{pmatrix}
\end{aligned}
\end{eqnarray}
Now defining the Rabi frequencies as,
\begin{eqnarray}
\begin{aligned}
     \Omega_{pr}&=\frac{\mathbf{d}_{14}\mathcal{\mathbf{E}}_{pr}}{\hbar}\\
      \Omega_{pu}&=\frac{\mathbf{d}_{24}\mathcal{\mathbf{E}}_{pu}}{\hbar}\\
       \Omega_{c}&=\frac{\mathbf{d}_{34}\mathcal{\mathbf{E}}_{c}}{\hbar}\\
\end{aligned}
   \end{eqnarray}
   Now the full Hamiltonian of the system is obtained as
   \begin{eqnarray}
       \mathcal{H}=\frac{\hbar}{2}\begin{pmatrix}
2\omega_{1}&0&0&-\Omega_{pr}e^{i\omega_{pr}t}\\
0&2\omega_{2}&0&-\Omega_{pu}e^{i\omega_{pu}t}\\
0&0&2\omega_{3}&-\Omega_{c}e^{i(\omega_{c}\pm \omega_{m})t}\\
-\Omega_{pr}e^{-i\omega_{pr}t}&-\Omega_{pu}e^{-i\omega_{pu}t}&-\Omega_{c}e^{-i(\omega_{c}\pm \omega_{m})t}&2\omega_{4}
\end{pmatrix}
   \end{eqnarray}
   To make this Hamiltonian time independent, we shifted to the corotating basis, and for this purpose, we define a unitary operator as
   \begin{eqnarray}
       \mathcal{\Tilde{U}}(t)=\begin{pmatrix}
e^{-i\omega_{pr}t}&0&0&0\\
0&e^{-i\omega_{pu}t}&0&0\\
0&0&e^{-i\omega_{c}t}&0\\
0&0&0&1
\end{pmatrix}
   \end{eqnarray}
   To ensure the Schrödinger equation holds in the corotating basis, the Hamiltonian obeys the following equation
   \begin{eqnarray}
       \mathcal{H}=i\hbar\frac{\partial \Tilde{\mathcal{U}}}{\partial t}\Tilde{\mathcal{U}}^{\dagger}+\Tilde{\mathcal{U}}\mathcal{H}\Tilde{\mathcal{U}}^{\dagger}
   \end{eqnarray}
  Thus, the total Hamiltonian in the corotating frame transforms into 
   \begin{eqnarray}
   \begin{aligned}
         \mathcal{H}=&\hbar\begin{pmatrix}
0&0&0&-\frac{\Omega_{pr}}{2}\\
0&(\Delta_{\textrm{pr}}-\Delta_{\textrm{pu}})&0&-\frac{\Omega_{pu}}{2}\\
0&0&(\Delta_{\textrm{pr}}-\Delta_{\textrm{c}}\pm \omega_{\textrm{m}})&-\frac{\Omega_{c}}{2}\\
-\frac{\Omega_{pr}}{2}&-\frac{\Omega_{pu}}{2}&-\frac{\Omega_{c}}{2}&\Delta_{\textrm{pr}}
\end{pmatrix}\\
=&\hbar\lbrace (\Delta_{\textrm{pr}}-\Delta_{\textrm{pu}})\sigma_{22}+(\Delta_{\textrm{pr}}-\Delta_{\textrm{c}}\pm \omega_{\textrm{m}})\sigma_{33}+\Delta_{\textrm{pr}}\sigma_{44}\rbrace -\frac{\hbar}{2}\left[\Omega_{\textrm{pr}}\sigma_{14}+\Omega_{\textrm{pu}}\sigma_{24}+\Omega_{\textrm{c}}\sigma_{34}+h.c\right]
   \end{aligned}
     \end{eqnarray}
$\sigma_{jk}=\vert j\rangle\langle k\vert$ is the transition operator for $j\neq k$. 

\section*{Appendix II}
The respective dephasing rates for the $\textrm{HE}_{\textrm{pu}}$ and $\textrm{HE}_{\textrm{c}}$ engines can be written as follows
\begin{eqnarray}\label{dephasing2}
\begin{aligned}
\gamma_{41}&=\Gamma_{41}+\Gamma_{42}+2 \textrm{R}_{14}+\textrm{R}_{24}\\ 
\gamma_{42}&=\Gamma_{41}+\Gamma_{42}+ \textrm{R}_{14}+2\textrm{R}_{24}\\
\gamma_{21}&= \textrm{R}_{14}+\textrm{R}_{24}
\end{aligned}
\end{eqnarray}
and 
\begin{eqnarray}\label{dephasing3}
\begin{aligned}
\gamma_{41}&=\Gamma_{41}+\Gamma_{43}+2 \textrm{R}_{14}+\textrm{R}_{34}\\ 
\gamma_{43}&=\Gamma_{41}+\Gamma_{43}+ \textrm{R}_{14}+2\textrm{R}_{34}\\
\gamma_{31}&= \textrm{R}_{14}+\textrm{R}_{34}
\end{aligned}
\end{eqnarray}
\section*{Appendix III}
 In Eqs. \eqref{mainsolution} and \eqref{mainsolution2}, the $G$ and $F$ are defined as
\begin{eqnarray}
    G=\frac{1}{2}\gamma_{14}
-i\Delta_{\textrm{pr}}+\frac{\Omega_{\textrm{pu}}^{2}}{4(\frac{1}{2}\gamma_{21}-i(\Delta_{\textrm{pr}}-\Delta_{\textrm{pu}}))}+\frac{(\frac{k_{c}\textrm{z}_{o}}{2})^{2}\Omega_{\textrm{c}}^{2}}{4(\frac{1}{2}\gamma_{31}
-i(\Delta_{\textrm{pr}}-\Delta_{\textrm{c}})\mp i\omega_{\textrm{m}})}
\end{eqnarray}
and
\begin{eqnarray}
    F=\frac{1}{2}\gamma_{43}
+i\Delta_{\textrm{c}}\mp i\omega_{\textrm{m}}+\frac{\Omega_{\textrm{pu}}^{2}}{4(\frac{1}{2}\gamma_{23}
+i(\Delta_{\textrm{pu}}-\Delta_{\textrm{c}})\mp i\omega_{\textrm{m}})}.
\end{eqnarray}
The population terms are explicitly defined as follows
 \begin{eqnarray}
  \begin{aligned}
    \rho_{11}&=\frac{XY(\Gamma_{41}+\textrm{R}_{14})}{4XY\textrm{R}_{14}+XY\Gamma_{41}+Y\textrm{R}_{14}\Gamma_{42}+X\textrm{R}_{14}\Gamma_{43}}  \\
     \rho_{22}&=\frac{Y\textrm{R}_{14}(\Gamma_{42}+X)}{4XY\textrm{R}_{14}+XY\Gamma_{41}+Y\textrm{R}_{14}\Gamma_{42}+X\textrm{R}_{14}\Gamma_{43}}\\
       \rho_{33}&=-\frac{-XY\textrm{R}_{14}-X\textrm{R}_{14}\Gamma_{43}}{4XY\textrm{R}_{14}+XY\Gamma_{41}+Y\textrm{R}_{14}\Gamma_{42}+X\textrm{R}_{14}\Gamma_{43}}\\
       \rho_{44}&=-\frac{XY\textrm{R}_{14}}{4XY\textrm{R}_{14}+XY\Gamma_{41}+Y\textrm{R}_{14}\Gamma_{42}+X\textrm{R}_{14}\Gamma_{43}}\
 \end{aligned}
 \end{eqnarray}
 where,
 \begin{eqnarray}
  \begin{aligned}
    Y&=\textrm{R}_{34}+\frac{\gamma_{43}(1+\frac{k_{0}z_{0}}{2})^{2}\Omega_{\textrm{c}}^{2}}{\gamma_{43}^{2}+4(\Delta_{\textrm{c}}\mp\omega_{\textrm{m}})^{2}}\\
    X&=\textrm{R}_{24}+\frac{\gamma_{42}\Omega_{\textrm{pu}}^{2}}{\gamma_{42}^{2}+4\Delta_{\textrm{pu}}^{2}}.
 \end{aligned}
 \end{eqnarray}

\section*{Acknowledgment}
The author expresses thanks to Jayanta Kr. Saha and Md. Mabud Hossain for their valuable discussions and critical reading of the paper. Jayanta Kr. Saha's thought-provoking questions regarding the theoretical model section assist the author in refining the mathematical expressions and enhancing the manuscript's readability. The author is also grateful to Arnab Ghosh for initially shedding light on the subject of the heat engine. Additionally, the author gratefully acknowledges the financial support provided for this research project by the West Bengal Government's Department of Higher Education, Science, Technology, and Bio-Technology (DHESTBT) under grant number 249(Sanc.)/ST/P/S \&T/16G-26/2017.  Furthermore, the author acknowledges the Ph.D. fellowship received under the Swami Vivekananda Merit-Means Scholarship (SVMCMS), offered by the Government of West Bengal, India, through Aliah University.

\newpage
\bibliographystyle{unsrt}
\bibliography{biblio}

\newpage
\begin{center}
\begin{scriptsize}
\begin{table}[!th]
\caption{The scaled temperature for the $\textrm{HE}_{\textrm{pu}}$ engine is limited to $\frac{\textrm{T}_{\textrm{B}}}{\textrm{T}_{0}} \leq 4$, with a reference temperature of $\textrm{T}_{0}=5000$K. The upper and lower limits of entropy (in in $\textrm{k}_{\textrm{B}}^{-1}$ unit) for $\textrm{HE}_{\textrm{pu}}$ engine are 6.110 and 1.527, respectively. The respective maximum scaled temperature ($\frac{\textrm{T}_{\textrm{max}}}{\textrm{T}_{0}}$), entropy ($\textrm{S}$) and emission rate ($\textrm{R}$) are obtained from equations \eqref{temp_max}, \eqref{entropy} and \eqref{emission} for the engine by integrating [eqs. \eqref{entropy} and \eqref{emission}] over the frequency range from -50 MHz to 50 MHz of emitted probe frequency $\omega_{\textrm{pr}}$. The other system parameters are $\textrm{T}_{41}=\textrm{T}_{42}$=5000 K, $\omega_{41}=4\times10^{15}$ (2$\pi$. Hz),  $\omega_{42}=3\times10^{15}$ (2$\pi$. Hz), $\Delta_{\textrm{pu}}=0$, and in calculation of $\frac{\textrm{T}_{\textrm{max}}}{\textrm{T}_{0}}$, the probe detuning is zero \textit{i.e.} $\Delta_{\textrm{pr}}=0$.}
\label{tab_pu}
\begin{center}
\begin{tabular}{ccccccc}
\hline
\hline
&&$\textrm{HE}_{\textrm{pu}}$ engine\\
\hline
Sl.&$\Omega_{\textrm{pu}}$&$\frac{\textrm{T}_{\textrm{max}}}{\textrm{T}_{0}}$&\textrm{S}&\textrm{R}\\
No.& (in $\gamma_{41}$)& &(in $\textrm{k}_{\textrm{B}}^{-1}$) &(in $s^{-1}$)\\
\hline
1&0& 1.438 &5.106&0.035 \\
2&0.5& 3.882 &3.283&1.542\\
3&1& 3.884 &3.246&1.597\\
4&1.5& 3.884 &3.239&1.608\\
5&2& 3.884 &3.237&1.612\\
6&2.5& 3.884 &3.236&1.613\\
\hline
\hline
\end{tabular}
\end{center} 
\end{table}
\end{scriptsize}
\end{center} 
\begin{center}
\begin{scriptsize}
\begin{table}[!th]
\caption{The scaled temperature for the $\textrm{HE}_{\textrm{c}}$ engine is limited to $\frac{\textrm{T}_{\textrm{B}}}{\textrm{T}_{0}} \leq 4$, with a reference temperature of $\textrm{T}_{0}=5000$K. The upper and lower limits of entropy (in $\textrm{k}_{\textrm{B}}^{-1}$ unit) for $\textrm{HE}_{\textrm{c}}$ engine are 6.110 and 1.527, respectively. The respective maximum scaled temperature ($\frac{\textrm{T}_{\textrm{max}}}{\textrm{T}_{0}}$), entropy ($\textrm{S}$) and emission rate ($\textrm{R}$) are obtained from equations \eqref{temp_max}, \eqref{entropy} and \eqref{emission} for the engine by integrating [eqs. \eqref{entropy} and \eqref{emission}] over the frequency range from -50 MHz to 50 MHz of emitted probe frequency $\omega_{\textrm{pr}}$. The other system parameters are $\textrm{T}_{41}=\textrm{T}_{43}$=5000 K, $\omega_{41}=4\times10^{15}$ (2$\pi$. Hz), $\omega_{43}=3\times10^{15}$ (2$\pi$. Hz), $\Delta_{\textrm{c}}=0$, and $k_{0}z_{0}=0.01$. In calculation of $\frac{\textrm{T}_{\textrm{max}}}{\textrm{T}_{0}}$, the probe detuning is zero \textit{i.e.} }$\Delta_{\textrm{pr}}=0$.\label{tab_c}
\begin{center}
\begin{tabular}{cccccc}
\hline
\hline
&&$\textrm{HE}_{\textrm{c}}$ engine\\
\hline
Sl.&$\omega_{\textrm{m}}$&$\Omega_{\textrm{c}}$&$\frac{\textrm{T}_{\textrm{max}}}{\textrm{T}_{0}}$&\textrm{S}&\textrm{R}\\
No.&in (2$\pi$. Hz)& (in $\gamma_{41}$)& &(in $\textrm{k}_{\textrm{B}}^{-1}$) &(in $s^{-1}$)\\
\hline
1&1&0& 1.438&5.106&0.035 \\
2&1&0.5&2.530 &5.155&0.033\\
3&1&1& 2.532&4.839&0.040\\
4&1&1.5& 2.535&3.926&0.100\\
5&1&2& 2.535&3.261&0.302\\
6&2&0& 1.438&5.106&0.035 \\
7&2&0.5& 2.183 &5.163&0.032\\
8&2&1& 2.185 &5.184&0.031\\
9&2&1.5& 2.186 &4.862&0.054\\
10&2&2& 2.186 &3.637&0.146\\
11&3&0& 1.438&5.106&0.035 \\
12&3&0.5&2.046  &5.181&0.032\\
13&3&1& 2.048 &5.186&0.028\\
14&3&1.5& 2.050 &5.128&0.037\\
15&3&2& 2.050 &5.062&0.085\\
\hline
\hline
\end{tabular}
\end{center} 
\end{table}
\end{scriptsize}
\end{center} 

\begin{center}
\begin{scriptsize}
\begin{table}[!th]
\caption{The scaled temperature for the $\textrm{HE}_{\textrm{pu, c}}$ engine is limited to $\frac{\textrm{T}_{\textrm{B}}}{\textrm{T}_{0}} \geq 2$, with a reference temperature of $\textrm{T}_{0}=5000$K. The upper and lower limits of entropy (in $\textrm{k}_{\textrm{B}}^{-1}$ unit) for $\textrm{HE}_{\textrm{pu,c}}$ engine are 6.110 and -3.05, respectively. The respective maximum scaled temperature ($\frac{\textrm{T}_{\textrm{max}}}{\textrm{T}_{0}}$), entropy ($\textrm{S}$) and emission rate ($\textrm{R}$) are obtained from equations \eqref{temp_max}, \eqref{entropy} and \eqref{emission} for the engine by integrating [eqs. \eqref{entropy} and \eqref{emission}] over the frequency range from -50 MHz to 50 MHz of emitted probe frequency $\omega_{\textrm{pr}}$. The other system parameters are $\Omega_{\textrm{pu}}=\gamma_{41}$, $\textrm{T}_{41}=\textrm{T}_{42}=\textrm{T}_{43}$=5000 K, $\omega_{41}=4\times10^{15}$ (2$\pi$. Hz), $\omega_{43}=\omega_{42}=3\times10^{15}$ (2$\pi$. Hz), $\Delta_{\textrm{pu}}=0$, $\Delta_{\textrm{c}}=0$ and $k_{0}z_{0}=0.01$. In calculation of $\frac{\textrm{T}_{\textrm{max}}}{\textrm{T}_{0}}$, the probe detuning is zero \textit{i.e.} }$\Delta_{\textrm{pr}}=0$.\label{tab_pu,c}
\begin{center}
\begin{tabular}{cccccc}
\hline
\hline
&&$\textrm{HE}_{\textrm{pu,c}}$ engine\\
\hline
Sl.&$\omega_{\textrm{m}}$&$\Omega_{\textrm{c}}$&$\frac{\textrm{T}_{\textrm{max}}}{\textrm{T}_{0}}$&\textrm{S}&\textrm{R}\\
No.&in (2$\pi$. Hz)& (in $\gamma_{41}$)& &(in $\textrm{k}_{\textrm{B}}^{-1}$) &(in $s^{-1}$)\\
\hline
1&1&0& 2.589&2.902&3.324 \\
2&1&0.5&2.812 &2.952&3.215\\
3&1&1& 2.824&2.916&3.309\\
4&1&1.5& 2.825 &2.915&3.310\\
5&1&2& 2.825&2.913&3.312\\
6&2&0& 2.589&2.902&3.324 \\
7&2&0.5& 3.014 &2.925&3.274\\
8&2&1& 3.017 &2.918&3.316\\
9&2&1.5& 3.018 &2.917&3.317\\
10&2&2& 3.018 &2.916&3.319\\
11&3&0& 2.589&2.902&3.324\\
12&3&0.5&2.975 &3.001&3.301\\
13&3&1& 3.187 &2.995&3.320\\
14&3&1.5& 3.188&2.995&3.320\\
15&3&2& 3.188 &2.994&3.321\\
\hline
\hline
\end{tabular}
\end{center} 
\end{table}
\end{scriptsize}
\end{center} 


\newpage


\end{document}